\documentclass[12pt]{article}
\usepackage{epsfig}
\usepackage{amssymb}
\usepackage{amsmath}
\usepackage{amsfonts}

\newcommand{\R}{\mathbb{R}}
\newcommand{\C}{\mathbb{C}}
\newcommand{\Z}{\mathbb{Z}}

\newcommand{\be}{\begin{equation}}
\newcommand{\ee}{\end{equation}}
\newcommand{\bea}{\begin{eqnarray}}
\newcommand{\eea}{\end{eqnarray}}
\newcommand{\nn}{\nonumber}
\newcommand{\kt}{\rangle}
\newcommand{\br}{\langle}

\newcommand{\ed}{\end{document}}

\newcommand{\rx}{{\rm x}}

\newcommand{\rp}{{\rm p}}

\newcommand{\rH}{{\rm H}}
\newcommand{\rX}{{\rm X}}
\newcommand{\rh}{{\rm h}}
\newcommand{\rE}{{\rm E}}
\newcommand{\sN}{\mbox{\tiny $N$}}
\newcommand{\approxN}{\stackrel{\sN}{\approx}}

\begin{document}

\title{Physical Aspects of Pseudo-Hermitian and $PT$-Symmetric
Quantum Mechanics}

\author{\\
Ali Mostafazadeh\thanks{E-mail address: amostafazadeh@ku.edu.tr}
~and Ahmet Batal\thanks{E-mail address: abatal@ku.edu.tr}\\
\\ Department of Mathematics, Ko\c{c} University,\\
Rumelifeneri Yolu, 34450 Sariyer,\\
Istanbul, Turkey}
\date{ }
\maketitle

\begin{abstract}
For a non-Hermitian Hamiltonian $H$ possessing a real spectrum, we
introduce a canonical orthonormal basis in which a previously
introduced unitary mapping of $H$ to a Hermitian Hamiltonian $h$
takes a simple form. We use this basis to construct the
observables $O_\alpha$ of the quantum mechanics based on $H$. In
particular, we introduce pseudo-Hermitian position and momentum
operators and a pseudo-Hermitian quantization scheme that relates
the latter to the ordinary classical position and momentum
observables. These allow us to address the problem of determining
the conserved probability density and the underlying classical
system for pseudo-Hermitian and in particular $PT$-symmetric
quantum systems. As a concrete example we construct the Hermitian
Hamiltonian $h$, the physical observables $O_\alpha$, the
localized states, and the conserved probability density for the
non-Hermitian $PT$-symmetric square well. We achieve this by
employing an appropriate perturbation scheme. For this system, we
conduct a comprehensive study of both the kinematical and
dynamical effects of the non-Hermiticity of the Hamiltonian on
various physical quantities. In particular, we show that these
effects are quantum mechanical in nature and diminish in the
classical limit. Our results provide an objective assessment of
the physical aspects of $PT$-symmetric quantum mechanics and
clarify its relationship with both the conventional quantum
mechanics and the classical mechanics.
\end{abstract}
\begin{center}
~~~~~PACS number: 03.65.-w
\end{center}
\vspace{2mm}

\tableofcontents



\section{Introduction}

Most of the recent publications on $PT$-symmetric quantum
mechanics focus on the study of the spectral properties of various
(non-Hermitian) $PT$-symmetric Hamiltonians. The results reported
in these publications are mainly mathematical. The purpose of the
present paper is to address some of the most basic problems
related to the physical aspects of $PT$-symmetric and more
generally pseudo-Hermitian quantum mechanics. In particular, we
will offer a complete description of the nature and the
construction of the physical observables and provide a method to
compute various physical quantities in these theories. We will
also elucidate the relationship between these theories and the
conventional classical and quantum mechanics.

As our approach is motivated by the mathematical results obtained
within the framework of the theory of pseudo-Hermitian operators
\cite{p1,p2,p3}, we begin our discussion by a brief review of the
relevant developments.

A central question that arises in connection with the current
interest in $PT$-symmetric quantum mechanics
\cite{bender-98,bender-99} is: ``What are the necessary and
sufficient conditions for the reality of the spectrum of a linear
operator?'' Ref.~\cite{p2} provides the following answer to this
question: If the operator acts in a Hilbert space ${\cal H}$ and
has a complete set of eigenvectors (i.e., it is diagonalizable)
then its spectrum is real if and only if (one and consequently all
of) the following equivalent conditions holds.
    \begin{itemize}
    \item[](C1) There exists a positive-definite operator\footnote{An
    operator is called positive-definite if it is Hermitian and has a strictly
    positive spectrum.} $\eta_+:{\cal H}\to{\cal H}$ that fulfils
        \be
        H^\dagger=\eta_+ H\eta_+^{-1},
        \label{ph}
        \ee
    i.e., $H$ is pseudo-Hermitian~\cite{p1} and the set\footnote{For
    a discussion of this set, see \cite{jmp-2004}.} of all
    the metric operators $\eta$ satisfying $H^\dagger=\eta H
    \eta^{-1}$ includes a positive-definite element.
    \item[](C2) $H$ is Hermitian with respect to some
    positive-definite inner product $\br\cdot,\cdot\kt_+$ on
    ${\cal H}$ (which is generally different from its
    defining inner product $\br\cdot|\cdot\kt$.) A specific
    choice for $\br\cdot,\cdot\kt_+$ is $\br\cdot|\eta_+\cdot\kt$.
    \item[](C3) $H$ may be mapped to a Hermitian Hamiltonian $h$
    by a similarity transformation, i.e., $H$ is quasi-Hermitian
    \cite{quasi,kresh}.
    \end{itemize}
The framework provided in Refs.~\cite{p1,p2} also explains the
connection with $PT$-symmetry. It turns out that, under the same
conditions, pseudo-Hermiticity of $H$ is equivalent to the
presence of an antilinear symmetry, $PT$-symmetry being the
primary example, \cite{p3,ss1}.

The condition that the Hamiltonian $H$ must have a complete set of
eigenvectors may be relaxed by extending the analysis of
\cite{p1,p2,p3} to block-diagonalizable linear operators as
discussed in \cite{p6,ss2}. However, note that physically this
condition is intertwined with the requirements of the quantum
measurement theory. The failure to satisfy it is equivalent to
allowing for the states that have zero overlap with all the energy
eigenstates. As a result, the total probability of measuring any
energy value for such a state is identically zero, i.e., one can
never perform an energy measurement on such a state; it must not
be possible to prepare it!

These physical considerations form the basis of a general
framework, called pseudo-Hermitian quantum mechanics
\cite{cjp-2003}, that allows for formulating a quantum theory
based on an eigenvalue problem for a linear operator $H$ acting in
a (complex) vector space $V$. A typical example is an eigenvalue
(Sturm-Liouville) equation for a differential operator acting in a
complex function space. Supposing that this eigenvalue problem has
a solution, i.e., there are eigenvectors $\psi_n\in V$, one lets
$V_H$ be the span of $\psi_n$, endows $V_H$ with an arbitrary
positive-definite inner product, Cauchy completes
\cite{reed-simon} this inner product space to a Hilbert space
${\cal H}$, and views $H$ as a (possibly densely defined) linear
operator acting in ${\cal H}$. Then, by construction, $H$ is a
diagonalizable operator acting in ${\cal H}$, and the results of
\cite{p1,p2,p3} apply.

As noted in \cite{p9}, the equivalence of the reality of the
spectrum of $H$ and the condition (C2) is the basic mathematical
result underlying the construction of the so-called $CPT$-inner
product for $PT$-symmetric quantum systems \cite{bbj}. Also as
shown in \cite{jpa-2003}, one can use the condition (C3) to map
$H$ to a Hermitian Hamiltonian $h$ acting in ${\cal H}$. If one
identifies the physical Hilbert space ${\cal H}_{\rm phys}$ of the
system with ${\cal H}$ endowed with the positive-definite inner
product $\br\cdot,\cdot\kt_+$, then $H$ and $h$ are unitarily
equivalent.

For models with a finite-dimensional Hilbert space the
construction of the Hermitian Hamiltonian $h$ is straightforward.
In some cases $h$ has a much simpler form than $H$,
\cite{jpa-2003}. The situation is quite different for systems with
an infinite-dimensional Hilbert space, as almost nothing specific
is known about the structure of $h$. It is nevertheless expected
to be a generally complicated nonlocal (non-differential) operator
\cite{jpa-2003}.

The study of systems with an infinite-dimensional Hilbert space is
particularly important, because it is for such systems that one
can seek for an underlying classical system and attempt to
formulate an associated quantization scheme. Obviously, a proper
treatment of these issues requires a careful study of the notion
of a physical observable in pseudo-Hermitian and, in particular,
$PT$-symmetric quantum mechanics.

It has recently been shown \cite{critique} that the formulation of
observables in $PT$-symmetric quantum mechanics as originally
proposed in \cite{bbj} and reiterated in \cite{bbj-ajp} is
inconsistent with its dynamical aspects and that enforcing the
rules of the standard measurement theory restricts the choice of
the observables $O_\alpha$ to linear operators that are Hermitian
with respect to the inner product of the physical Hilbert space
${\cal H}_{\rm phys}$.\footnote{To resolve the inconsistency
reported in \cite{critique}, the authors of \cite{bbj} have
recently revised their definition of observables
\cite{bbj-erratum}. As noted in \cite{comment}, it is not known if
this corrected definition is generally compatible with the
requirements of the quantum measurement theory. When the contour
defining the boundary conditions of the problem is the real line,
the definition reduces to ours (and consequently the above
compatibility condition holds). But even in this case it is a more
restrictive definition as it implies that the Hamiltonian must be
(not only $PT$-symmetric but also) symmetric, i.e., in
$x$-representation it is a symmetric (infinite) matrix. This leads
to some undesirable consequences \cite{comment}.} Accepting this
definition for the observables, one can easily show that the
unitary mapping that maps $H$ to $h$ also maps the observables
$O_\alpha$ to the Hermitian operators $o_\alpha$ acting in ${\cal
H}$. This in turn means that a physical system described by the
Hilbert space ${\cal H}_{\rm phys}$, the $PT$-symmetric
Hamiltonian $H$, and the observables $O_\alpha$ may be
equivalently described by the Hilbert space ${\cal H}$, the
Hermitian Hamiltonian $h$, and the observables $o_\alpha$.

In this paper we will introduce a canonical basis in which the
construction of the Hermitian Hamiltonian $h$ and the physical
observables $O_\alpha$ simplifies considerably. This allows us to
determine the underlying classical system and develop a
pseudo-Hermitian quantization scheme. We will also introduce and
construct the pseudo-Hermitian position operator, the
corresponding position wave functions, and the conserved
probability density. As a concrete application of our general
results we perform a thorough investigation of the $PT$-symmetric
square well Hamiltonian, computing the corresponding Hermitian
Hamiltonian $h$, the observables $O_\alpha$  (in particular the
pseudo-Hermitian position operator), the probability density, the
position expectation values, and the localized states. We will
also describe the effects of the non-Hermiticity of the
Hamiltonian on the latter quantities and discuss the underlying
classical Hamiltonian.

Throughout this paper we will assume that $H$ is diagonalizable
and has a nondegenerate, real, discrete spectrum. The extension of
the results to degenerate spectra may be easily achieved following
the approach of \cite{p1,p2,p3,p9}. The presence of a continuous
part of the spectrum does not lead to any serious complications
either. For example see \cite{cqg,rqm1}.

\section{Canonical Metric Basis}

Let ${\cal H}$ be a Hilbert space and $H:{\cal H}\to{\cal H}$ be a
diagonalizable linear (Hamiltonian) operator having a real,
nondegenerate, discrete spectrum. Following \cite{p1,p2,p3,p9}, we
shall label the eigenvalues of $H$ with $E_n$ and let
$\{|\psi_n\kt\}$ denote a basis of ${\cal H}$ consisting of the
eigenvectors $|\psi_n\kt$ of $H$,
    \be
    H|\psi_n\kt=E_n|\psi_n\kt.
    \label{eg-va-H}
    \ee
Then one can construct another basis $\{|\phi_n\kt\}$ of ${\cal
H}$ that satisfies \cite{p3}
    \be
    H^\dagger|\phi_n\kt=E_n|\phi_n\kt,~~~~~~~~
    \br\phi_n|\psi_m\kt=\delta_{mn},~~~~~~~~
    \sum_n|\psi_n\kt\br\phi_n|=1.
    \label{bi-ortho}
    \ee
In particular, $\{|\psi_n\kt,|\phi_n\kt\}$ form a biorthonormal
system \cite{bi-ortho}, and
    \be
    H=\sum_n E_n|\psi_n\kt\br\phi_n|,~~~~~~~~~
    H^\dagger=\sum_n E_n|\phi_n\kt\br\psi_n|.
    \label{resol}
    \ee
Here and throughout this paper, for any linear operator acting in
${\cal H}$, $A^\dagger$ stands for the adjoint of $A$, i.e., the
unique linear operator satisfying $\br\cdot|A\cdot\kt=\br
A^\dagger\cdot|\cdot\kt$.

A central result of \cite{p1} is that the operator
    \be
    \eta_+=\sum_n|\phi_n\kt\br\phi_n|
    \label{eta+}
    \ee
satisfies (\ref{ph}). It is also manifestly positive-definite,
because it satisfies $\eta_+=w^\dagger w$, where $w:=\sum_n
|n\kt\br\phi_n|$ and $\{|n\kt\}$ is an orthonormal basis of ${\cal
H}$, and that it is invertible, with the inverse given by
    \be
    \eta_+^{-1}=\sum_n|\psi_n\kt\br\psi_n|.
    \label{eta+inv}
    \ee

We can use (\ref{eta+}) to introduce the positive-definite inner
product:
    \be
    \br\cdot,\cdot\kt_+:=\br\cdot|\eta_+\cdot\kt,
    \label{inn-prod}
    \ee
and identify the physical Hilbert space ${\cal H}_{\rm phys}$ with
the underlying vector space of ${\cal H}$ endowed with this inner
product. This means that as complex vector spaces ${\cal H}$ and
${\cal H}_{\rm phys}$ are identical, but as Hilbert spaces they
are not.

In view of (\ref{ph}), the Hermitian Hamiltonian $h$ of condition
(C3) has the form \cite{jpa-2003}:
    \be
    h=\rho\,H\,\rho^{-1},
    \label{h=}
    \ee
where $\rho:=\sqrt\eta_+$ is the unique positive(-definite) square
root of $\eta_+$. The transformation $H\to h$ corresponds to the
linear mapping $|\psi\kt\to\rho|\psi\kt$. It is a simple exercise
to check that, for any pair $|\psi\kt,|\psi'\kt$ of state vectors:
$\br\psi,\psi'\kt_+=\br\psi|\eta_+\psi'\kt=\br\rho\psi|\rho\psi'\kt.$
Hence as a mapping of ${\cal H}_{\rm phys}$ onto ${\cal H}$,
$\rho$ is a unitary operator.\footnote{A linear map $U:{\cal
H}_1\to{\cal H}_2$ between two inner product (in particular
Hilbert) spaces ${\cal H}_1$ and ${\cal H}_2$ with inner products
$\br\cdot,\cdot\kt_1$ and $\br\cdot,\cdot\kt_2$ is said to be a
unitary operator if for all $\zeta,\chi\in{\cal H}_1$, we have
$\br U(\zeta),U(\chi)\kt_2= \br \zeta,\chi\kt_1$,
\cite{reed-simon}. $U$ is unitary if and only if it is invertible
(one-to-one and onto) and $U^{-1}=U^\dagger$.}

Now, consider a physical system $S$ that is described by the
Hilbert space ${\cal H}_{\rm phys}$, the Hamiltonian $H$, and the
observables $O_\alpha$ that are Hermitian operators acting in
${\cal H}_{\rm phys}$.\footnote{Being a Hermitian operator acting
in ${\cal H}_{\rm phys}$, the Hamiltonian $H$ is also an
observable. But as operators acting in ${\cal H}$ neither $H$ nor
$O_\alpha$ are Hermitian.} Because $\rho:{\cal H}_{\rm
phys}\to{\cal H}$ is a unitary transformation, $O_\alpha:{\cal
H}_{\rm phys}\to{\cal H}_{\rm phys}$ is Hermitian if and only if
$o_\alpha:=\rho\,O_\alpha\rho^{-1}:{\cal H}\to{\cal H}$ is
Hermitian. This, in particular, means that the observables
$O_\alpha$ may be constructed from the Hermitian operators
$o_\alpha$ according to \cite{critique}
    \be
    O_\alpha=\rho^{-1}o_\alpha\rho.
    \label{observable=}
    \ee
Consequently, we can also describe the physical system $S$ using
the original Hilbert space ${\cal H}$, the Hermitian Hamiltonian
$h$, and the observables $o_\alpha$. The two descriptions are
physically identical as there is a one-to-one correspondence
between the states and the observables used in these descriptions
and more importantly the physical quantities such as the
transition amplitudes or expectation values of the observables do
not depend on the choice of the description.

The main ingredient of the above construction is the operator
$\rho=\sqrt\eta_+$. It has three important properties:
\begin{itemize}
\item[P1.] As an operator mapping ${\cal H}_{\rm phys}$ to ${\cal
H}$, it is a unitary operator;

\item[P2.] As an operator mapping ${\cal H}$ to ${\cal H}$, it is
a Hermitian operator;

\item[P3.] As an operator mapping ${\cal H}_{\rm phys}$ to ${\cal
H}_{\rm phys}$, it is also a Hermitian operator.\footnote{This can
be easily checked: $\br\cdot,\rho\cdot\kt_{+}=
\br\cdot,|\eta_+\rho\cdot\kt=\br\cdot|\rho\eta_+\cdot\kt=
\br\rho\cdot|\eta_+\cdot\kt=\br\rho\cdot,\cdot\kt_{+}$.} In
particular, both $\rho$ and $\eta_+=\rho^2$ are physical
observables.
\end{itemize}
Property P2 suggests that a natural method for computing the
operators $h$ and $O_\alpha$ is to use an orthonormal basis
$\{|\epsilon_n\kt\}$ of ${\cal H}$ that consists of the
eigenvectors\footnote{Here we suppress the degeneracy labels for
the eigenvectors $|\epsilon_n\kt$ for simplicity. Note also that
in view of the non-uniqueness \cite{p4,p9} of $\eta_+$ one can
assume without loss of generality that the eigenvalues of $\eta_+$
are nondegenerate.} $|\epsilon_n\kt$ of $\eta_+$. Denoting the
eigenvalues of $\eta_+$ by $\epsilon_n$, we have
    \be
    \eta_+|\epsilon_n\kt=\epsilon_n|\epsilon_n\kt,
    ~~~~~~\br\epsilon_m|\epsilon_n\kt=\delta_{mn},
    ~~~~~~\sum_n|\epsilon_n\kt\br\epsilon_n|=1.
    \label{epsilon}
    \ee
These in turn imply
    \be
    \eta_+=\sum_n \epsilon_n |\epsilon_n\kt\br \epsilon_n|,
    ~~~~~~~~
    \rho=\sum_n \sqrt{\epsilon_n}~ |\epsilon_n\kt\br \epsilon_n|.
    \label{resol-2}
    \ee
In the following we shall refer to $\{|\epsilon_n\kt\}$ as a {\em
canonical metric basis}.

Let $A$ be a linear operator acting in ${\cal H}$, we can uniquely
identify $A$ with its matrix representation $(A_{mn})$ in the
basis $\{|\epsilon_n\kt\}$, where
    \be
    A_{mn}:=\br\epsilon_m|A|\epsilon_n\kt.
    \label{matrix}
    \ee
Because $\{|\epsilon_n\kt\}$ is an orthonormal basis of ${\cal
H}$, the matrix elements of $A^\dagger$ are given by
    \be
    A^\dagger_{mn}=A_{nm}^*.
    \label{matrix-dagger}
    \ee
In particular, $A$ is Hermitian with respect to the defining inner
product $\br\cdot|\cdot\kt$ of ${\cal H}$ if and only if
$(A_{mn})$ is a Hermitian (possibly infinite) matrix, i.e.,
$A_{mn}^*=A_{nm}$.

The following important identities follow from (\ref{ph}),
(\ref{eta+}), (\ref{eta+inv}) and (\ref{h=}).
    \bea
    &&\epsilon_n^{-1}\epsilon_m H_{mn}=H^\dagger_{mn}=H_{nm}^*,
    \label{matrix-ph}\\
    &&\epsilon_n=\sum_m|\br\phi_m|\epsilon_n\kt|^2=
    \left(\sum_m|\br\psi_m|\epsilon_n\kt|^2\right)^{-1},
    \label{matrix-eps}\\
    && h_{mn}=\sqrt{\frac{\epsilon_m}{\epsilon_n}}\; H_{mn}.
    \label{matrix-h=}
    \eea
Furthermore, let $o:{\cal H}\to{\cal H}$ be a Hermitian operator
and $O:=\rho^{-1}o\rho$, then
    \be
    O_{mn}=\sqrt{\frac{\epsilon_n}{\epsilon_m}}\; o_{mn}.
    \label{matrix-o}
    \ee
Eqs.~(\ref{matrix-h=}) and (\ref{matrix-o}) provide the following
expressions for the Hermitian Hamiltonian $h:{\cal H}\to{\cal H}$
and the observables $O:{\cal H}_{\rm phys}\to{\cal H}_{\rm phys}$.
    \bea
    h&=&\sum_{m,n}\sqrt{\frac{\epsilon_m}{\epsilon_n}}\; H_{mn}\;
    |\epsilon_m\kt\br\epsilon_n|,
    \label{h=2}\\
    O&=&\sum_{m,n}\sqrt{\frac{\epsilon_n}{\epsilon_m}}\; o_{mn}\;
    |\epsilon_m\kt\br\epsilon_n|.
    \label{O=2}
    \eea

\section{Classical System and Its Pseudo-Hermitian Canonical
Quantization}

For ${\cal H}=L^2(\R)$, we can define the {\em
$\eta_+$-pseudo-Hermitian position ($X$) and momentum ($P$)
operators} according to
    \be
    X:=\sum_{m,n}\sqrt{\frac{\epsilon_n}{\epsilon_m}}\; x_{mn}\;
    |\epsilon_m\kt\br\epsilon_n|,~~~~~~~~~
    P:=\sum_{m,n}\sqrt{\frac{\epsilon_n}{\epsilon_m}}\; p_{mn}\;
    |\epsilon_m\kt\br\epsilon_n|,
    \label{position}
    \ee
where $x_{mn}:=\br\epsilon_m|x|\epsilon_n\kt$,
$p_{mn}:=\br\epsilon_m|p|\epsilon_n\kt$, and $x$ and $p$ are the
usual position and momentum operators acting in ${\cal
H}=L^2(\R)$.

Clearly, the $\eta_+$-pseudo-Hermitian position and momentum
operators satisfy the canonical commutation relation
    \be
    [X,P]=i\hbar\,1.
    \label{ccr}
    \ee
Indeed, together with the identity operator $1$, they provide a
unitary irreducible representation of the Weyl-Heisenberg algebra
which has the physical Hilbert space ${\cal H}_{\rm phys}$ as the
representation space. The fact that by construction this
representation is unitarily equivalent to the standard
representation of the Weyl-Heisenberg algebra (that has ${\cal
H}=L^2(\R)$ as the representation space) is a manifestation of
von-Neumann's celebrated uniqueness theorem.\footnote{This theorem
states that up to unitary equivalence the Weyl-Heisenberg algebra
has a unique unitary irreducible (projective) representation
\cite{reed-simon}.}

Having introduced the $\eta_+$-pseudo-Hermitian position and
momentum operators, we can also speak of the following {\em
$\eta_+$-pseudo-Hermitian canonical quantization} of classical
systems:
    \be
    x_c\to X,~~~~~~p_c\to P,~~~~~~~\{\cdot,\cdot\}_c\to
    -i\hbar^{-1}[\cdot,\cdot],
    \label{p-cq}
    \ee
where $x_c, p_c,$ and $\{\cdot,\cdot\}_c$ stand for classical
position, momentum, and Poisson bracket, respectively. For
instance, $\eta_+$-pseudo-Hermitian quantization of the classical
Hamiltonian for a free particle leads to the pseudo-Hermitian
quantum Hamiltonian:
    \be
    H_{\rm free}=\frac{P^2}{2m}=\rho^{-1}
    \left[\frac{p^2}{2m}\right]\rho,
    \label{free}
    \ee
which is a generally nonlocal (non-differential) operator.

Note that in general the Hamiltonian operator $H$, that is used to
construct the metric operator $\eta_+$ and consequently define the
above notion of pseudo-Hermitian quantization, does not have the
standard form $P^2/(2m)+V(X)$. For example, a $PT$-symmetric
Hamiltonian of the standard form \cite{bender-98} (with a
complex-valued potential $v(x)$),
    \be
    H=\frac{p^2}{2m}+v(x)=
    \rho\left[\frac{P^2}{2m}+v(X)\right]\rho^{-1},
    \label{non-standard}
    \ee
cannot generally be expressed in the form $P^2/2m+V(X)$ for any
real-valued function $V$. Nevertheless, because (in light of
property P3) $\rho$ is also a physical observable, one can express
$\rho$ and $\rho^{-1}$ and consequently the Hamiltonian
(\ref{non-standard}) as certain power series in $X$ and $P$
(modulo commutation relations (\ref{ccr}).) This in turn implies
that the classical Hamiltonian $H_c$, whose
$\eta_+$-pseudo-Hermitian quantization yields $H$, is not
generally of the standard (Kinetic+Potential) type. Rather it is a
complicated (non-polynomial) function of $x_c$ and $p_c$.

The classical Hamiltonian $H_c$ may also be obtained using the
Hermitian Hamiltonian $h$ which according to (\ref{h=}) and
(\ref{non-standard}) takes the form
    \be
    h=\rho\left[\frac{p^2}{2m}+v(x)\right]\rho^{-1}.
    \label{PT-h}
    \ee
Again this is a nonlocal operator which can be expressed as a
power series in $p$ with $x$-dependent coefficients. This is
because (according to property P2) $\rho$ and $\rho^{-1}$ are
Hermitian operators acting in ${\cal H}$. The classical
Hamiltonian may be obtained by replacing $x$ and $p$ in the
expression for $h$ by their classical counterparts $x_c$ and
$p_c$, respectively. Clearly the resulting $H_c$ is identical with
the one obtained from $H$.

Next, we wish to recall a simple procedure for associating a power
series in $x$ and $p$ (i.e., a pseudo-differential operator) to a
nonlocal linear operator $K:L^2(\R)\to L^2(\R)$. Suppose that $K$
may be expressed in terms of its kernel ${\cal K}:\R^2\to\C$
according to
    \be
    (K\psi)(x)=\int_\R {\cal K}(x,x')\psi(x')dx'.
    \label{kernel}
    \ee
Then for real analytic wave functions, that form a dense subset of
$L^2(\R)$, we can expand $\psi(x')$ appearing on the right-hand
side of (\ref{kernel}) in Taylor series about $x$. Substituting
the result in (\ref{kernel}), we find $(K\psi)(x)=\hat K\psi(x)$
where
    \bea
    \hat K&:=&\sum_{\ell=0}^\infty (-i\hbar)^\ell
    a_\ell(x)\frac{d^\ell}{dx^\ell},\label{kernel-2}\\
    a_\ell(x)&=&\frac{i^\ell}{\ell!\hbar^\ell}\int_\R K(x,x')(x'-x)^\ell dx'.
    \label{moments}
    \eea
As a result, we have the following (densely defined) identity
    \be
    K=\sum_{\ell=0}^\infty a_\ell(x)\,p^\ell.
    \label{kernel-3}
    \ee
If the operator $K$ is Hermitian, we can express (\ref{kernel-3})
in a manifestly Hermitian form, namely
    \be
    K=\frac{1}{2}
    \sum_{\ell=0}^\infty [a_\ell(x)\,p^\ell+p^\ell a_\ell(x)^*].
    \label{kernel-30}
    \ee
The classical counterpart of this operator is the following
real-valued function of the phase space ($\R^2$).
    \be
    K_c(x_c,p_c):=\sum_{\ell=0}^\infty
    \Re[a_\ell(x_c)]\,p_c^\ell,~~~~~~~~~
    x_c,p_c\in\R,
    \label{classical}
    \ee
where $\Re$ means `Real part of'.

The results reported in this section clearly generalize to the
Hilbert spaces ${\cal H}(V)$ where $V$ is $\R^n$ or a
topologically equivalent subset of $\R^n$. Together with the
results of the preceding section, they lead to the following
prescription for determining the classical Hamiltonian for a
pseudo-Hermitian (particularly $PT$-symmetric) quantum system:
    \begin{enumerate}
    \item Given the Hamiltonian $H$, compute a metric operator
    $\eta_+$;
    \item Diagonalize $\eta_+$ and construct the corresponding
    canonical metric basis $\{|\epsilon_n\kt\}$;
    \item Compute the matrix elements $H_{mn}$ of $H$ in this
    basis and use (\ref{h=2}) to obtain the Hermitian Hamiltonian
    $h$;
    \item Apply the above described method of associating a
    pseudo-differential operator to the operator $h$, express the
    latter in a manifestly Hermitian form $(h+h^\dagger)/2$, and
    take $x\to x_c$ and $p\to p_c$ in the resulting expression.
    This yields a classical Hamiltonian $H_c$ for the theory.
    \end{enumerate}
The Hamiltonian $H_c$ obtained in this way generally involve
$\hbar$. The strictly classical Hamiltonian will correspond to
evaluating $\hbar\to 0$ limit of $H_c$. The latter is an
admissible prescription only if this limit exists.

We end this section by making the last step of the above
prescription more specific. Using (\ref{h=2}), (\ref{moments}),
(\ref{kernel-3}) and (\ref{classical}), identifying the kernel of
$h$ with $\br x|h|x'\kt$, and denoting the normalized
eigenfunctions of $\eta_+$ by $\varepsilon_n$, i.e.,
    \be
    \varepsilon_n(x)=\br x|\epsilon_n\kt,
    \label{epsilon-x}
    \ee
we have
    \bea
    h&=&\sum_{\ell=0}^\infty a_\ell(x)\;p^\ell,~~~~~~~~~~
    H_c=\sum_{\ell=0}^\infty \Re[a_\ell(x_c)]\;p_c^\ell,
    \label{h=3}\\
    a_\ell(x)&=&\frac{i^\ell}{\ell! \hbar^\ell}
    \sum_{m,n}\sqrt{\frac{\epsilon_m}{\epsilon_n}}\;
    H_{mn}\; \varepsilon_m(x)
    \int_\R \varepsilon_n(x')^* (x'-x)^\ell dx'.
    \label{a=}
    \eea
Admittedly, the computation of $h$ as outlined above is too
complicated to be done exactly. In Sec.~5, we study its
application to a simple $PT$-symmetric model where a particularly
useful approximation scheme allows for computing $h$ with any
desired accuracy. Finally, we should like to add that the above
prescription for computing $h$ may also be used to compute the
pseudo-Hermitian observables such as the position operator $X$.

\section{Localized States, Position Wave Functions, and
the Probability Density}

Having introduced the $\eta_+$-pseudo-Hermitian position operator
$X$ we can identify its (generalized \cite{bohm-qm}) eigenvectors
$\xi^{(x)}$ with the localized states of the system. They are
defined by
    \be
    X \xi^{(x)}=x\,\xi^{(x)},~~~~~~~~\forall x\in\R.
    \label{xi}
    \ee
In view of the identity $X=\rho^{-1}x\rho$, we have
    \be
    \xi^{(x)}=\rho^{-1}|x\kt,
    \label{xi=}
    \ee
where $|x\kt$ are the usual position kets satisfying, for all
$x,x'\in\R$,
    \be
    \br x|x'\kt=\delta(x-x'),~~~~~~~~\int_\R dx\,|x\kt\br x|=1.
    \label{ortho-complete-x}
    \ee
Using these relations and the fact that $\rho:{\cal H}_{\rm
phys}\to{\cal H}$ is a unitary mapping, we can establish the
orthonormality and completeness relations for the localized states
$\xi^{(x)}$:
    \be
    \br\xi^{(x)},\xi^{(x')}\kt_+=\delta(x-x'),~~~~~~~~
    \int_\R dx\;\Xi^{(x)}=1,
    \label{ortho-complete-xi}
    \ee
where $\Xi^{(x)}$ denotes the projection operator defined by
    \be
    \Xi^{(x)}\psi:=\br\xi^{(x)},\psi\kt_+\xi^{(x)},~~~~~~~
    {\rm for~all}~~~~~~~\psi\in{\cal H}_{\rm phys}.
    \label{Xi}
    \ee

Next, consider a particle\footnote{Here by a particle we mean a
quantum system having $\R$ as its classical configuration space.}
whose state at a fixed time $t_0$ is described by the state vector
$\psi\in{\cal H}_{\rm phys}$. We can introduce the {\em position
wave function}:
    \be
    \Psi(x):=\br\xi^{(x)},\psi\kt_+=\br x|\rho|\psi\kt,
    \label{position-wf}
    \ee
and use (\ref{ortho-complete-xi}) to expand the state vector $\psi$
in the position basis $\{\xi^{(x)}\}$ according to
    \be
    \psi=\int_\R \Psi(x)\xi^{(x)} dx.
    \label{psi=wf}
    \ee
As seen from (\ref{position-wf}), the position wave function
$\Psi(x)$ is generally different from $\psi(x)$. This is a direct
consequence of the fact that $H$ as an operator acting in ${\cal
H}=L^2(\R)$ fails to be Hermitian. Furthermore, in view of
(\ref{ortho-complete-xi}) and (\ref{position-wf}),
    \[\parallel\Psi\parallel^2:=\int_\R |\Psi(x)|^2 dx=
    \int_\R |\br\xi^{(x)},\psi\kt_+|^2 dx=\int_\R
    \br\psi,\Xi^{(x)}\psi\kt_+dx=\br\psi,\psi\kt_+<\infty.\]
Hence as a function mapping $\R$ to $\C$, the wave function $\Psi$
belongs to $L^2(\R)$.\footnote{Here we view $\Psi$ as an abstract
vector belonging to $L^2(\R)$. The state vector $\psi$ also
belongs to the Hilbert space ${\cal H}$ which coincides with
$L^2(\R)$. However, these two copies of $L^2(\R)$ should not be
confused.} The converse is also true in the sense that every
square-integrable function $\Psi$ defines a state vector
$\psi\in{\cal H}_{\rm phys}$. Therefore, we may identify $L^2(\R)$
with the vector space of position wave functions $\Psi$ for the
system. It is also a straightforward exercise to show that the
assignment $F$ of a wave function $\Psi=:F(\psi)$ to each state
vector $\psi$, viewed as a map $F:{\cal H}_{\rm phys}\to L^2(\R)$,
is a unitary operator. In order to see this, let
$\psi,\phi\in{\cal H}_{\rm phys}$ be arbitrary state vectors and
$\Psi=F(\psi)$ and $\Phi=F(\psi)$, then
    \[\br F(\psi)|F(\phi)\kt=
    \br\Psi|\Phi\kt=\int_\R\Psi(x)^*\Phi(x)dx=
    \int_\R\br\psi,\xi^{(x)}\kt_+\br\xi^{(x)},\phi\kt_+dx=
    \int_\R\br\psi,\Xi^{(x)}\phi\kt_+dx=
    \br\psi,\phi\kt_+.\]

In the following we will assume without loss of generality that
$\psi$ is normalized with respect to the inner product
$\br\cdot,\cdot\kt_+$, i.e., set
$\br\psi,\psi\kt_+=\parallel\Psi\parallel^2=1$.

According to the standard quantum measurement theory, the
probability of finding the particle in a region $V\subseteq\R$ at
time $t_0$ is given by
    \be
    \Pi_V(\psi):=\int_V |\br\xi^{(x)},\psi\kt_+|^2dx.
    \label{prob}
    \ee
Hence
    \be
    \varrho(x):=|\br\xi^{(x)},\psi\kt_+|^2=|\Psi(x)|^2
    \label{prob-dens}
    \ee
is the {\em probability density} of the localization of the
particle in space.

Unlike the naive ``probability density''
$\varrho^{(0)}(x):=|\psi(x)|^2$, $\varrho(x)$ defines a conserved
total probability. This follows from the fact that $H$ is
Hermitian with respect to the inner product $\br\cdot,\cdot\kt_+$
of ${\cal H}_{\rm phys}$. It is instructive to demonstrate the
conservation of total probability in the position representation.
In order to do so, consider the time-evolution of the state vector
$\psi$ as determined by the Schr\"odinger equation:
    \be
    i\hbar\frac{d}{dt}\psi(t)=H\psi(t).
    \label{sch-eq}
    \ee
Computing the inner product of both sides of this equation with
$\xi^{(x)}$ (using the inner product $\br\cdot,\cdot\kt_+$) and
employing the completeness relation given in
(\ref{ortho-complete-xi}), we find
    \be
    i\hbar\frac{d}{dt}\Psi(x;t)=\hat h\Psi(x;t),
    \label{sch-eq-wf}
    \ee
where $\Psi(x;t):=\br\xi^{(x)},\psi(t)\kt_+$ and $\hat
h:L^2(\R)\to L^2(\R)$ is defined by
    \be
    \hat h\Psi(x;t):=\int_\R {\cal K}(x,y)\Psi(y;t)\,dy,~~~~~~
    {\cal K}(x,y):=\br\xi^{(x)},H\xi^{(y)}\kt_+.
    \label{sch-eq-kernel}
    \ee
Because, as an operator acting in ${\cal H}_{\rm phys}$, $H$ is
Hermitian,
    \[ {\cal K}(x,y)^*=\br\xi^{(x)},H\xi^{(y)}\kt_+^*=
    \br H\xi^{(x)},\xi^{(y)}\kt_+^*=\br\xi^{(y)},H\xi^{(x)}\kt_+
    =\hat{\cal K}(y,x).\]
This is sufficient to conclude that $\hat h$ is a Hermitian
operator acting in $L^2(\R)$. As a result, in the position
representation the dynamics is determined by a Hermitian
Hamiltonian; the time-evolution operator, $e^{-i(t-t_0)\hat
h/\hbar}$, for the position wave functions is unitary; and the
total probability
    \[\Pi_\R(\psi(t))=\int_\R |\Psi(x;t)|^2 dx\]
is conserved.

The Hamiltonian operator $\hat h$ is directly related to the
Hermitian Hamiltonian $h$. Substituting (\ref{xi=}) in
(\ref{sch-eq-kernel}) and using (\ref{inn-prod}) and (\ref{h=}),
we have
    \bea
    {\cal K}(x,y)&=&\br x| \rho^{-1}\eta_+H\rho^{-1}|y\kt=
    \br x|h|y\kt,\nn\\
    \hat h\Psi(x;t)&=&\int_\R \br x|h|y\kt\Psi(y;t)\,dy=
    \int_\R \br x|h|y\kt\br y|\rho\psi(t)\kt\,dy=
    \br x|h\rho\psi(t)\kt.\nn
    \eea
Hence, in light of (\ref{position-wf}), the Hamiltonian $\hat h$ is
the usual position representation of the Hermitian Hamiltonian $h$,
i.e.,
    \be
    \br x|h=\hat h\br x|.
    \label{h-hat}
    \ee

This relationship between the Hamiltonian operators $\hat h$ and
$h$ extends to all the physical observables. Given an observable
$O$ acting in ${\cal H}_{\rm phys}$ and the corresponding operator
$o=\rho O\rho^{-1}$ acting in ${\cal H}$, we can define an
associated Hermitian operator $\hat o$ acting $L^2(\R)$ that
realizes the action of $O$ on a state vector $\psi\in{\cal H}_{\rm
phys}$ in terms of the corresponding position wave function $\Psi$
according to
    \be
    O\psi=\int_\R [\hat o\Psi(x)]\,\xi^{(x)}dx.
    \label{O=hat-O}
    \ee
The operator $\hat o$ is the position representation of the
abstract operator $o$;
    \be
    \br x|o=\hat o\br x|.
    \label{o-hat}
    \ee
In view of (\ref{O=hat-O}), (\ref{o-hat}),
(\ref{ortho-complete-xi}), and (\ref{position-wf}), the
expectation value of $O$ in a state described by the normalized
state vector $\psi$ and position wave function $\Psi$ is given by
    \be
    \br\psi,O\psi\kt_+=\int_{-\infty}^\infty
    \Psi(x)^*\,\hat o\Psi(x)\:dx.
    \label{exp-val}
    \ee

As shown in the preceding paragraphs, one can formulate both the
dynamics and the kinematics of the theory using the position wave
functions $\Psi$. In this formulation the observables and in
particular the Hamiltonian are Hermitian operators acting in
${\cal H}$ similarly to the conventional quantum mechanics. In
order to use this formulation, however, one needs a more explicit
expression for the wave function $\Psi$. We may derive such an
expression using the canonical metric basis $\{|\epsilon_n\kt\}$.
In view of, (\ref{epsilon}), (\ref{resol-2}), (\ref{epsilon-x}),
and (\ref{position-wf}), we have
    \be
    \Psi(x)=\sum_n f_n\;\varepsilon_n(x),~~~~~~~~
    f_n:=\epsilon_n^{1/2}\int_\R \varepsilon_n(x')^*\psi(x')\,dx'.
    \label{wf=epsilon}
    \ee

\section{Application to the $PT$-Symmetric Square Well}

The $PT$-symmetric square well potential, originally introduced by
Znojil in \cite{z1}, provides a simple model with generic
properties of general $PT$-symmetric potentials. Its Hamiltonian
is given by
    \be
    H=\frac{{p}^2}{2m}+v({x}),
    \label{sw}
    \ee
where
    \be
    v({x})=\left\{\begin{array}{ccc}
    \infty&{\rm for}& { x}\notin (-\frac{L}{2},\frac{L}{2})\\
    i \zeta&{\rm for}& { x}\in(-\frac{L}{2},0)\\
    -i \zeta&{\rm for}& { x}\in(0,\frac{L}{2}),
    \end{array}\right.
    \label{pwsw-1}
    \ee
$L\in\R^+$ and $\zeta\in\R$. Usually one employs units in which
$L=2$, $m=1/2$, and $\hbar=1$. This is equivalent to using the
dimensionless variables
    \be
    x\to{\rm x}:=\left(\frac{2}{L}\right)\,x,~~~~
    p\to{\rm p}:=\left(\frac{L}{2\hbar}\right)\,p,~~~~
    \zeta\to Z:=\left(\frac{mL^2}{2\hbar^2}\right)\,\zeta,
    \label{rescale}
    \ee
and working with the dimensionless Hamiltonian:
    \be
    \rH:=\left(\frac{mL^2}{2\hbar^2}\right)\, H=
    \rp^2+{\rm v(x)},
    \label{scaled-H}
    \ee
where
    \be
    {\rm v(x)}=
    \left\{\begin{array}{ccc}
    \infty&{\rm for}& {\rm x}\notin (-1,1)\\
    i  Z&{\rm for}& {\rm x}\in(-1,0)\\
    -i Z&{\rm for}& {\rm x}\in(0,1).
    \end{array}\right.
    \label{pwsw}
    \ee

In the ${\rm x}$-representation, the eigenvalue problem for $\rH$
takes the form\footnote{The eigenvalues $E_n$ of the Hamiltonian
(\ref{sw}) are given by $2\hbar^2 \rE_n/(mL^2)$.}
    \be
    \left[-\frac{d^2}{d{\rm x}^2}+{\rm v}({\rm x})-\rE_n\right]
    \psi_n(\rx)=0.
    \label{eg-va}
    \ee
The Hilbert space to which the eigenvectors $\psi_n$ belong
is\footnote{The Hilbert space associated with the unscaled
Hamiltonian $H$ is obtained by changing $1$ in (\ref{Hilbert}) to
$L/2$.}
    \be
    {\cal H}=\left\{\psi\in L^2(\R)~|~\psi(x)=0~{\rm for}~|x|\geq 1
    \right\}=
    \left\{\psi\in L^2([-1,1])~|~\psi(\pm 1)=0\right\}.
    \label{Hilbert}
    \ee
Clearly, $\rH$ is not Hermitian with respect to the defining inner
product $\br\cdot|\cdot\kt$ of ${\cal H}$. This is an indication
that ${\cal H}$ is not the physical Hilbert space ${\cal H}_{\rm
phys}$. In order to specify the latter we should determine an
appropriate metric operator $\eta_+$. This in turn requires the
solution of the eigenvalue equation (\ref{eg-va}).

The eigenvalue problem for the $PT$-symmetric square well admits
an essentially explicit solution. A detailed discussion is
provided in \cite{z1,bmq}. If $Z$ is below the critical value
$Z_\star\approx 4.48$ the Hamiltonian has a real spectrum
\cite{z1}.\footnote{$Z=Z_\star$ marks an exceptional point
\cite{exceptional} where two real eigenvalues cross in such a way
that the Hamiltonian becomes non-diagonalizable. Once $Z$ exceeds
$Z_\star$, $H$ regains its diagonalizability, but a pair of
complex-conjugate eigenvalues appear in its spectrum
\cite{p6,jmp-2004}. Increasing the value of $Z$ indefinitely one
encounters an infinite number of exceptional points passing each
of which produces a complex-conjugate pair of eigenvalues.} For
these values of the `non-Hermiticity' parameter $Z$, the Hilbert
spaces ${\cal H}$ and ${\cal H}_{\rm phys}$ are identical as
complex vector spaces, i.e., they are obtained by endowing their
common vector space with different inner products.

In this paper we will only be concerned with the case $0\leq
Z<Z_\star$. For these values of $Z$, one obtains the following
complete set of eigenfunctions of $H$:
    \bea
    \psi_n(\rx)&=&\left\{\begin{array}{ccc}
    \psi_{n-}(\rx)&{\rm for}& \rx\in[-1,0]\\
    \psi_{n+}(\rx)&{\rm for}& \rx\in[0,1],\end{array}\right.
    \label{eg-fu1}\\ \nn\\
    \psi_{n\pm}(\rx)&:=&\mbox{\Large $
    \frac{\alpha_n\sinh[\kappa_{n\pm}(1\mp x)]}{
    \sinh(\kappa_{n\pm})}$},
    \label{eg-fu2}
    \eea
where $\alpha_n$ are arbitrary nonzero real coefficients,
    \bea
    \kappa_{n\pm}&=&s_n\mp it_n,
    \label{kappa}\\
    s_n&:=&\frac{Z}{2t_n},
    \label{sn}
    \eea
and $t_n$ with $n\in\Z^+$ are the real solutions of the
transcendental equation:
    \be
    (Z/t_n)\sinh\left(Z/t_n\right)+2t_n\,\sin(2t_n)=0.
    \label{tn}
    \ee
The eigenvalues $\rE_n$ are given by
    \be
    \rE_n=-(\kappa_{n+}^2+iZ)=t_n^2-s_n^2.
    \label{en}
    \ee

Usually the coefficients $\alpha_n$ are fixed arbitrarily
\cite{z1} or kept as unimportant free coefficients \cite{bmq}. We
will fix them in such a way that in the limit $Z\to 0$, the
eigenfunctions $\psi_n$ of (\ref{eg-fu1}) tend to the well-known
normalized eigenfunctions $\psi_n^{(0)}$ of the conventional
(Hermitian) square well Hamiltonian (the case $Z=0$):
    \be
    \lim_{Z\to 0}\psi_n=\psi_n^{(0)}:=\left.\psi_n\right|_{_{Z=0}}.
    \label{conti}
    \ee

Because, by construction, $\psi_n$ are also the eigenfunctions of
the $PT$ operator, the continuity requirement (\ref{conti})
constrains $\psi_n^{(0)}$ to be $PT$-invariant. The normalized and
$PT$-invariant eigenfunctions of the Hermitian infinite square
well potential ($Z=0$) are, up to a sign, given by
    \be
    \psi_n^{(0)}(\rx)=i^{\mu_n}
    \sin\left[\frac{\pi n}{2}\,(\rx+1)\right],
    \label{z=0}
    \ee
where
    \be
    \mu_n:=\frac{1+(-1)^n}{2}.
    \label{sign}
    \ee
The eigenfunctions (\ref{z=0}) form an orthonormal basis of the
Hilbert space (\ref{Hilbert}). We will denote the corresponding
abstract basis vectors by $|n\kt$, i.e.,
    \be
    \br\rx|n\kt:=\psi_n^{(0)}(\rx),~~~~~~\forall n\in\Z^+.
    \label{n}
    \ee

The continuity requirement (\ref{conti}) together with
Eq.~(\ref{z=0}) restrict the coefficients $\alpha_n$ of the
eigenfunctions (\ref{eg-fu1}). Specifically, if we only keep the
leading order term in powers of $Z$ and neglect the higher order
terms, we find
    \be
    \alpha_n=(-1)^{\lfloor\frac{n}{2}\rfloor}\left(\frac{Z}{\pi
    n}\right)^{\mu_n},
    \label{alpha=}
    \ee
where $\lfloor\frac{n}{2}\rfloor$ stands for the integer part of
$\frac{n}{2}$. Both the eigenvalues $\rE_n$ and the eigenfunctions
$\psi_n$ are therefore determined once we obtain the solutions of
(\ref{tn}). As shown in \cite{z1}, this equation may be easily
solved numerically for various values of $Z$. In this paper, we
will solve this equation perturbatively by expanding the relevant
quantities in powers of $Z$.

\subsection{Perturbative Calculation of $\psi_n$ and $E_n$}

Suppose that $t_n$ admits a power series expansion about $Z=0$:
    \be
    t_n=\sum_{k=0}^\infty t_n^{(k)} Z^k,
    \label{t-exp}
    \ee
where $t_n^{(k)}\in\R$ are to be determined. Substituting
(\ref{t-exp}) in (\ref{tn}), expanding both sides of the resulting
equation in powers of $Z$, solving it term by term for
$t_n^{(k)}$, and using (\ref{t-exp}), (\ref{sn}) and (\ref{en}),
we find
    \bea
    t_n&=&\left(\frac{\pi n}{2}\right)\left\{1-(-1)^{n}\nu^2
    -\left[3+\frac{(-1)^n\pi^2n^2}{6}\right]\nu^4+{\cal O}(\nu^6)
    \right\},
    \label{tn=}\\
    s_n&=&\left(\frac{\pi n\nu}{2}\right)\left\{1+(-1)^{n}\nu^2+
    \left[4+\frac{(-1)^n\pi^2n^2}{6}\right]\nu^4+
    {\cal O}(\nu^6)\right\},
    \label{sn=}\\
    \rE_n&=&\left(\frac{\pi n}{2}\right)^2\left\{1
    -\left[1+2(-1)^n\right]\nu^2-
    \left[5+2(-1)^n+\frac{(-1)^n\pi^2n^2}{3}\right]\nu^4+
    {\cal O}(\nu^6)\right\},
    \label{en=}
    \eea
where
    \be
    \nu:=\frac{2Z}{(\pi n)^2},
    \label{nu}
    \ee
and ${\cal O}(\nu^k)$ stands for terms of order $\nu^k$ and
higher.

Eqs.~(\ref{tn=}) -- (\ref{en=}) reveal the curious fact that the
effective perturbation parameter is $2Z/(\pi n)^2$.\footnote{The
condition that the above perturbative calculations would be
unreliable for the ground state, i.e., $\nu\approx 1$, corresponds
to $Z\approx 4.92$ which is slightly above the critical value
$Z_\star=4.48$.} This is a clear indication that the
non-Hermiticity of the Hamiltonian $\rH$ only affects the low
lying energy levels. This property of the $PT$-symmetric square
well Hamiltonian --- which has been previously known \cite{z1} ---
is particularly significant, for as we explain below it implies
that within the confines of the perturbation theory all the
infinite sums appearing in the expressions (\ref{eta+}),
(\ref{h=2}), and (\ref{wf=epsilon}) for the metric operator
$\eta_+$, the Hermitian Hamiltonian $\rh$, and the position wave
functions may be safely truncated. For example, for $Z=1$,
$\nu\approx 0.2/n^2$. Therefore, if we set
$\rE_n^{(0)}:=\rE_n|_{\nu=0}=(\pi n/2)^2$ and use
$q_n:=|\rE_n/\rE_n^{(0)}-1|$ as a measure of the contribution of
the non-Hermiticity of the Hamiltonian to the energy eigenvalues,
we find for $n>10$: $\nu<2\times 10^{-3}$ and $q_n<1.3\times
10^{-5}$.\footnote{The value $Z=1$, say for an electron ($m\approx
10^{-30}$ Kg) confined in a nanometer size well ($L\approx 10^{-9}
m$), corresponds to an energy scale $\zeta\approx 0.1~{\rm ev}$
for the potential (\ref{pwsw-1}). This is comparable with the
ground state energy ($E_1\approx 0.5~{\rm ev}$) of the
corresponding Hermitian infinite square well potential.} More
generally, we can ignore the effects of the non-Hermiticity
parameter $Z$ for all the computations involving the levels with
$n>10$ and still obtain results that are accurate at least up to
three decimal places.

In the following, we will employ an approximation scheme that
neglects the effects of the non-Hermiticity parameter $Z$ for all
levels with $n$ greater than a given number $N$. In view of the
above discussion, the results obtained using this approximation
will have an accuracy of the order of
    \be
    \nu_N=\frac{2Z}{(\pi N)^2}.
    \label{nuN}
    \ee
We will respectively refer to $N$ and $\nu_N$ as the `order' and
the `accuracy index' of our approximation scheme.

\subsection{Construction of a Canonical Metric Basis}

Having obtained the expression for (\ref{tn=}) and (\ref{sn=}) for
$t_n$ and $s_n$, we can compute $\kappa_n$ and use
Eqs.~(\ref{eg-fu1}), (\ref{eg-fu2}), and (\ref{alpha=}) to
determine the eigenfunctions $\psi_n$ of the Hamiltonian $\rH$ as a
power series in $\nu$.

The computation of a metric operator $\eta_+$, however, involves
the eigenfunctions $\phi_n$ of the adjoint $\rH^\dagger$ of $\rH$. It
is easy to see that $\rH^\dagger=\rH|_{_{Z\to-Z}}$. This suggests that
    \be
    \chi_n:=\psi_n|_{_{Z\to-Z}}
    \label{chi}
    \ee
are eigenfunctions of $\rH^\dagger$. The eigenfunctions $\phi_n$,
that together with $\psi_n$ form a biorthonormal system for the
Hilbert space ${\cal H}$, are obtained by properly normalizing
$\chi_n$. They are given by
    \be
    \phi_n(\rx)=N_n^{-1}\:\chi_n(\rx),
    \label{phi=}
    \ee
where
    \bea
    N_n&:=&\br\psi_n|\chi_n\kt=\int_{-1}^1\psi_n(\rx)^*\chi_n(\rx)\,
    d\rx\nn\\
    &=&
    \frac{2\alpha_{n+}\alpha_{n-}\left[
    1-\cos(2t_n)\cosh(2s_n)+
    \frac{t_n\sin(2t_n)[\cos(2t_n)-\cosh(2s_n)]}{s_n^2+t_n^2}\right]}{
    [\cos(2t_n)-\cosh(2s_n)]^{2}},
    \label{Nn}
    \eea
and
    \be
    \alpha_{n+}:=\alpha_n,~~~~~~~~~
    \alpha_{n-}:=\left.\alpha_n\right|_{_{Z\to-Z}}.
    \label{anpm=}
    \ee

Next, we construct the metric operator (\ref{eta+}) using the
approximation scheme described in the preceding section and the
orthonormal basis $\{|n\kt\}$ consisting of the eigenvectors
(\ref{n}) of the ordinary Hermitian infinite square well.

In the $N$-th order approximation, we have
    \be
    |\psi_n\kt\approxN |\phi_n\kt\approxN |n\kt,~~~~~~~~{\rm
    for~all}~n>N,
    \label{Nth-order}
    \ee
where `~$\approxN$~' stands for an equality that is valid up to
terms of order $\nu_N$. Combining (\ref{Nth-order}) with
(\ref{eta+}) we have
    \bea
    \eta_+&\approxN&\sum_{n=1}^N|\phi_n\kt\br\phi_n|+\sum_{n=N+1}^\infty
    |n\kt\br n|=1+\delta\eta_+^{(\sN)},
    \label{eta+=3}\\
    \delta\eta_+^{(\sN)}&:=&\sum_{n=1}^N\left(|\phi_n\kt\br\phi_n|-
    |n\kt\br n|\right).
    \label{delta-eta}
    \eea
This relation shows that the metric operator $\eta_+$ is
essentially determined by its projection onto the span ${\cal
H}_N$ of $|n\kt$ with $n\leq N$. Indeed, at this order of
approximation, ${\cal H}_N$ may also be identified with the span
of $\psi_n$ with $n\leq N$, or the span of $\phi_n$ with $n\leq
N$.

Clearly, the Hilbert space ${\cal H}$ is the direct sum of ${\cal
H}_N$ and its orthogonal complement ${\cal
H}_N^\perp:=\{\zeta\in{\cal
H}|\br\zeta|\psi\kt=0,\forall\psi\in{\cal H}_N\}$. As shown by
(\ref{eta+=3}) both of these are invariant subspaces
\cite{linear-algebra} of $\eta_+$. Therefore, one can solve the
eigenvalue problem for $\eta_+$ by restricting it onto ${\cal
H}_N$ and ${\cal H}_N^\perp$ and diagonalize the resulting
operators separately. The restriction of $\eta_+$ onto ${\cal
H}_N^\perp$ coincides with that of the identity operator. In
particular, it is diagonalized in the basis $\{|n\kt: n>N\}$, and
we have
    \be
    |\epsilon_n\kt\approxN|n\kt~~~{\rm and}~~~\epsilon_n\approxN1,
    ~~~~~~{\rm
    for~all}~n>N.
    \label{Nth-order2}
    \ee

The restriction of $\eta_+$ onto ${\cal H}_N$ yields a Hermitian
operator having a Hermitian matrix representation ${\cal E}$ in
the basis $\{|n\kt: n\leq N\}$. According to (\ref{eta+=3}), the
matrix elements of ${\cal E}$ are given by
    \be
    {\cal E}_{mn}=\br m|\eta_+|n\kt=\sum_{k=1}^N\br m|\phi_k\kt
    \br\phi_k|n\kt,
    \label{E}
    \ee
which in view of (\ref{chi}) -- (\ref{anpm=}), (\ref{eg-fu1}) --
(\ref{sn}), (\ref{alpha=}), (\ref{tn=}), and (\ref{sn=}) can be
computed explicitly. The computation of the metric basis vectors
$|\epsilon_n\kt$ with $n\leq N$ is equivalent to the
diagonalization of the Hermitian matrix ${\cal E}$. The latter can
be done both numerically and perturbatively.

Let $\{\vec e_n\}$ be a set of orthonormal eigenvectors of ${\cal
E}$ so that ${\cal E}\vec e_n=e_n \vec e_n$. Then clearly, up to
permutations of the labels, $e_n$ coincide with $\epsilon_n$ for
$n\leq N$. The canonical metric basis vectors $|\epsilon_n\kt$,
with $n\leq N$, are also related to the eigenvectors $\vec e_n$.
To make this relation explicit, we introduce the unitary $N\times
N$ matrix ${\cal U}$ whose columns coincide with the vectors $\vec
e_n$. Then, in view of (\ref{resol-2}), it is a straightforward
exercise to show that
    \be
    |\epsilon_n\kt\approxN \sum_{m=1}^N {\cal
    U}_{mn}|m\kt~~~{\rm and}~~~\epsilon_n\approxN e_n,~~~~~~{\rm
    for~all}~n\leq N.
    \label{epsilon=}
    \ee

Clearly, the above approximation scheme would be consistent only
if in the above calculation of ${\cal E}_{mn}$, $e_n$, and $\vec
e_n$ one takes into account the contribution of the terms of order
$\nu^\ell$ for which $\nu^\ell\geq\nu_{N}$. We can use
Eqs.~(\ref{nu}) and (\ref{nuN}) to make this condition more
explicit. A simple calculation shows that the negligible terms are
those of order $\nu^\ell$ with $\ell>\ell_n$, where
    \be
    \ell_n:=\frac{\ln (N)+r}{\ln
    n+r},~~~~r:=\ln\left(\frac{\pi}{\sqrt{2 Z}}\right)\approx
    0.789-\frac{\ln Z}{2}.
    \label{bound}
    \ee
Clearly, for smaller values of $n$ one should include higher order
corrections, the highest order term being of the order
$\nu^{\ell_1}$. For example, for $Z\leq 1$ and $N\leq 25$,
$\ell_1\leq 5.03$ and one can safely ignore ${\cal O}(\nu^6)$. For
the values $Z=1$ and $N=25$, the results will have a minimum
accuracy of the order of $\nu_N\approx 3.2\times 10^{-4}$.
Similarly for $Z\leq 0.5$ and $N\leq 100$, $\ell_1\leq 5.02$ and
one can again ignore ${\cal O}(\nu^6)$. The minimum accuracy
corresponding to the values $Z=0.5$ and $N=100$ is of the order of
$\nu_N\approx 1.0\times 10^{-5}$. Note that for all the above
values of $Z$ and $N$ the terms given explicitly in
Eqs.~(\ref{tn=}) -- (\ref{en=}) are sufficient to perform a
consistent perturbative calculation. A direct check of the
validity of this statement is to compare the exact value
$\rE_n^{\rm exact}$ of $\rE_n$ obtained by an accurate numerical
solution of (\ref{tn}) and the perturbative value $\rE_n^{\rm
pert.}$ of $\rE_n$ calculated using (\ref{en=}) and ignoring
${\cal O}(\nu^6)$. Clearly, the largest difference is for $n=1$.
If we express
    \[ \rE_1^{\rm exact}=\frac{\pi^2}{4}\,(1+\epsilon_1^{\rm
    exact}),~~~~~~~~
    \rE_1^{\rm pert.}=\frac{\pi^2}{4}\,(1+\epsilon_1^{\rm
    pert.}),\]
we find, for $Z=1$, $\epsilon_1^{\rm exact}=0.0415652$,
$\epsilon_1^{\rm pert.}=0.0415527$. This is in complete agreement
with our expectations, because the difference, $\epsilon_1^{\rm
exact}-\epsilon_1^{\rm pert.}= 1.24465\times 10^{-5}$, is much
smaller that the accuracy index $\nu_{N=25}\approx 3.2\times
10^{-4}$.

Next, observe that our approximation, in particular
(\ref{eta+=3}), corresponds to $\eta_+\approxN \eta_+^{(N)}$ where
    \[\eta_+^{(N)}\psi:=\left\{\begin{array}{ccc}
    \eta_+\psi&{\rm if}&\psi\in{\cal H}_N\\
    \psi&{\rm if}&\psi\in{\cal H}_N^\perp.\end{array}\right.\]
It is a reliable approximation, only if the distance between
$\eta_+^{(N)}$ and the identity operator $1$, as defined by
    \be
    \sigma_N:=\sqrt {{\rm trace}[(\eta_+^{(N)}-1)^2]}=
    \sqrt {{\rm trace}[(\delta\eta_+^{(\sN)})^2]},
    \label{spec1}
    \ee
has a finite large $N$-limit. This makes $\sigma_N$ a useful
measure of the validity of the above approximate calculation of
the canonical metric basis $\{|\epsilon_n\kt\}$. Clearly,
$\sigma_N$ is just the Frobenius or Euclidean distance
\cite{Horn-Johnson} between ${\cal E}$ and the $N\times N$
identity matrix $I$:
    \be
    \sigma_N=\parallel {\cal E}-I \parallel_2=
    \sqrt {{\rm trace}[({\cal E}-I)^2]}=
    \left[\sum_{n=1}^N (\epsilon_n-1)^2\right]^{1/2},
    \label{spec}
    \ee
where $\parallel\cdot\parallel_2$ stands for the Frobenius
(Euclidean) norm \cite{Horn-Johnson}, defined for every square
matrix $M$ by
    \be
    \parallel M\parallel_2:=\sqrt{{\rm trace}(M^\dagger M)}.
    \label{spec-norm}
    \ee

\begin{figure}[ht]
    \vspace{1cm}
    \begin{center}
    \epsffile{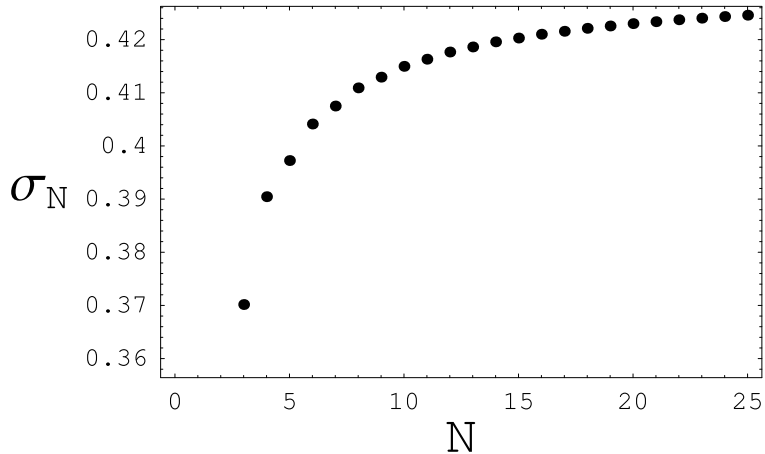}
    \end{center}
    \centerline{
    \parbox{12cm}{\caption{Graph of $\sigma_N$ as a function of $N$,
    for $Z=1$.}\label{fig1}}}
    \end{figure}

Fig.~\ref{fig1} shows a plot of $\sigma_N$ for $Z=1$ and $N\leq
25$ for which we can neglect ${\cal O}(\nu^6)$ in our calculations
while retaining the consistency of our approximation scheme. The
graph of $\sigma_N$ clearly shows the desired behavior even for
$N\approx 20$. The effective slope: $\sigma_{N}-\sigma_{N-1}$ of
the graph has the values $4.4\times 10^{-4}$ and $2.5\times
10^{-4}$ for N=20 and N=25, respectively.

Next, we describe another way of checking the reliability of our
approximation scheme. In view of (\ref{Nth-order}) and
(\ref{resol}), the matrix elements
    \be
    \rH_{mn}^{(0)}:=\br m|\rH|n\kt
    \label{H-mn-zero}
    \ee
may be approximated as
    \be
    \rH_{mn}^{(0)}\approxN \left\{\begin{array}{ccc}
    {\cal X}^{(\sN)}_{mn}&{\rm for}& m,n\leq N\\
    \frac{1}{4}\:\pi^2 n^2\,\delta_{mn}&{\rm for}& m,n> N,
    \end{array}\right.
    \label{H-check}
    \ee
where
    \be
    {\cal X}^{(\sN)}_{mn}:=\sum_{k=1}^N \rE_k\br
    m|\psi_k\kt\br\phi_k|n\kt
    \label{H-check-1}
    \ee
are computed by substituting (\ref{tn=}) and (\ref{sn=}) in
(\ref{kappa}) and using~(\ref{eg-fu2}), (\ref{en=}), and
(\ref{phi=}).

\begin{figure}[ht]
    \centerline{\epsffile{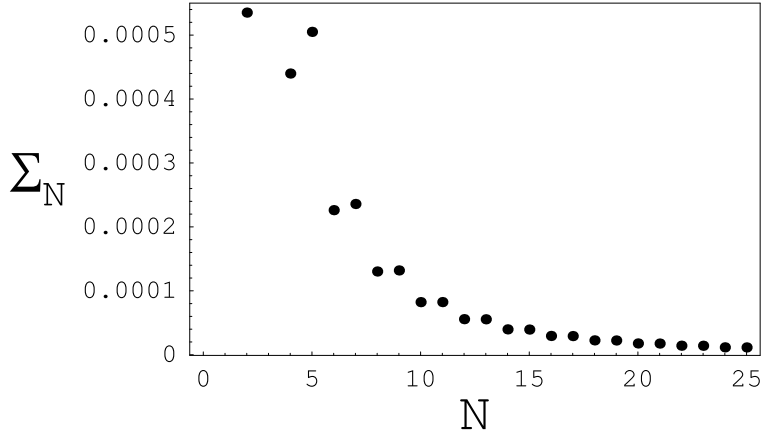}}
    \centerline{
    \parbox{12cm}{
    \caption{Graph of $\Sigma_N$ of Eq.~(\ref{Sigma-N})
    as a function of $N$, for $Z=1$. Note that
    $\Sigma_{25}\approx 1.1\times 10^{-5}$.}\label{fig2}}}
    \end{figure}

Consider the $N\times N$ matrix ${\cal X}^{(\sN)}$ with entries
(\ref{H-check-1}) and let ${\cal Y}^{(\sN)}$ be the $N\times N$
matrix with entries ${\cal Y}^{(\sN)}_{mn}:=\rH_{mn}^{(0)}$ for
all $m,n\leq N$. Then ${\cal Y}^{(\sN)}$ may be computed exactly
using (\ref{scaled-H}), (\ref{z=0}), (\ref{n}), and
(\ref{H-mn-zero}), whereas the calculation of ${\cal X}^{(\sN)}$
uses our approximation scheme. In order to compare ${\cal
X}^{(\sN)}$ and ${\cal Y}^{(\sN)}$, we will first introduce the
normalized matrices:
    \[\hat{\cal X}^{(\sN)}:=\frac{{\cal X}^{(\sN)}}{\parallel
    {\cal X}^{(\sN)}\parallel_2},~~~~~
    \hat{\cal Y}^{(\sN)}:=\frac{{\cal Y}^{(\sN)}}{\parallel
    {\cal Y}^{(\sN)}\parallel_2},\]
and use the Euclidean distance between the matrices $\hat {\cal
X}^{(\sN)}$ and $\hat {\cal Y}^{(\sN)}$, namely
    \be
    \Sigma_N:=\parallel \hat{\cal X}^{(\sN)}-\hat{\cal Y}^{(\sN)}
    \parallel_2,
    \label{Sigma-N}
    \ee
as a measure of the accuracy of our approximation. A reliable
approximation corresponds to a vanishing large-$N$ limit of
$\Sigma_N$.\footnote{Our use of the normalized matrices $\hat{\cal
X}^{(\sN)}$ and $\hat{\cal Y}^{(\sN)}$ stem from the fact that $H$
is not a bounded operator. It allows for the interpretation of the
term `a vanishing large-$N$ limit' as `$\Sigma_N\ll 1$ for
sufficiently large $N$'. In practice this means $\Sigma_N \leq
\nu_N$.} Fig.~\ref{fig2} shows the plot of $\Sigma_N$ for $N\leq
25$ and $Z=1$.
Table~\ref{tab1} lists the values of $\Sigma_N$ for various values
of $N$ and the corresponding values for the accuracy index
$\nu_N$. The results indicate that even for $N=10$ we have a
highly reliable approximation.

    \begin{table}[h]
    \vspace{.5cm}
    \begin{center}
  \begin{tabular}{||c||c|c|c||}
  \hline \hline
  $N$ & $\Sigma_N$ & $S_N$ & $\nu_N$ \\
  \hline\hline
  $10$ & $8.2\times 10^{-5}$& $1.7\times 10^{-5}$& $2.0\times 10^{-3}$\\
  \hline
  $15$ & $3.9\times 10^{-5}$& $9.0\times 10^{-6}$& $9.0\times 10^{-4}$\\
  \hline
  $20$ & $1.7\times 10^{-5}$& $3.4\times 10^{-6}$ & $5.1\times 10^{-4}$\\
  \hline
  $25$ & $1.1\times 10^{-5}$& $2.4\times 10^{-6}$ & $3.2\times 10^{-4}$\\
  \hline \hline\end{tabular}
  \end{center}
    \centerline{
    \parbox{12cm}{
    \caption{Values of $\Sigma_N$ of Eq.~(\ref{Sigma-N}), $S_n$
    of Eq.~(\ref{S-N}), and the accuracy index $\nu_N=2Z/(\pi N)^2$
    for $Z=1$ and various relevant values of $N$.}\label{tab1}}}
    \end{table}

\subsection{Construction of the Hermitian Hamiltonian $h$}

We can use the above approximation scheme to compute the Hermitian
Hamiltonian $\rh:=\rho\rH\rho^{-1}$ (respectively $h=\rho
H\rho^{-1}=2\hbar^2\rh/(mL^2)$ ) that is associated with the
$PT$-symmetric square well Hamiltonian $\rH$ (respectively $H$).
In order to do this we first use (\ref{h=2}), (\ref{Nth-order}),
(\ref{Nth-order2}), (\ref{epsilon=}), to express $\rh$ in the form
    \bea
    \rh&\approxN& \sum_{m,n=1}^N\sqrt{\frac{\epsilon_m}{\epsilon_n}}
    \;\rH_{mn}|\epsilon_m\kt\br\epsilon_n|+
    \sum_{m,n=N+1}^\infty \rH_{mn}^{(0)}|m\kt\br n|+\nn\\
    &&\sum_{m=1}^N\sum_{n=N+1}^\infty
    \left[\sqrt\epsilon_m H_{mn}|\epsilon_m\kt\br n|+
    {\frac{1}{\sqrt\epsilon_m}}\, H_{nm}
    |n\kt\br\epsilon_m|\right]
    =\rH+\delta \rH,
    \label{h=H+H}
    \eea
where $\rH_{mn}^{(0)}$ is given by (\ref{H-mn-zero}),
    \bea
     \rH_{mn}&=&\br\epsilon_m|\rH|\epsilon_n\kt\approxN
     \left\{\begin{array}{ccc}
    \sum_{j,k=1}^N {\cal U}_{mj}^\dagger \rH_{jk}^{(0)}\,
    {\cal U}_{kn}&~~~&\forall m,n\leq N,\\
    \br\epsilon_m|\rH|n\kt\approxN
    \br\epsilon_m|\rH|\psi_n\kt=\rE_n
    \br\epsilon_m|n\kt\approxN 0&~~~&\forall m\leq N,~n>N,
    \end{array}\right.
    \label{H-mn=}\\
    \delta \rH&:=&\sum_{m,n=1}^N \left(
    \sqrt{\frac{\epsilon_m}{\epsilon_n}}\;
    \rH_{mn}|\epsilon_m\kt\br\epsilon_n|-
    \rH_{mn}^{(0)}|m\kt\br n|\right).
    \label{delta-H}
    \eea
Substituting (\ref{H-mn=}) in (\ref{delta-H}) and using
(\ref{epsilon=}) and
    \be
    {\cal E}_{jk}^{\pm 1/2}=\sum_{m=1}^N {\cal U}_{jm}
    \:\epsilon_m^{\pm 1/2}\:{\cal U}^\dagger_{mk},
    \label{root}
    \ee
we find
    \be
    \delta \rH=\sum_{m,n=1}^N \delta \rH_{mn}^{(0)}|m\kt\br n|,
    \label{delta-H-2}
    \ee
where, for all $m,n\in\{1,2,\cdots,N\}$,
    \be
    \delta \rH_{mn}^{(0)}:=\sum_{j,k=1}^N {\cal E}^{1/2}_{mj}\,\rH_{jk}^{(0)}\,
    {\cal E}^{-1/2}_{kn} -\rH^{(0)}_{mn}.
    \label{tilde-H}
    \ee

\begin{figure}[h]
    \centerline{\epsffile{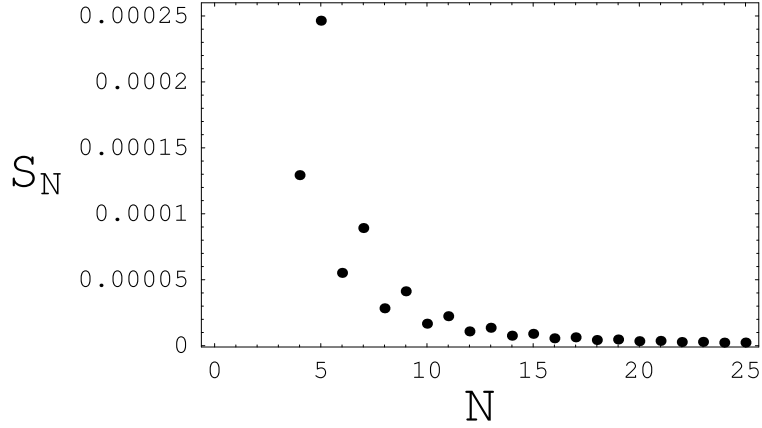}}
    \centerline{
    \parbox{12cm}{
    \caption{Graph of $S_N$ of Eq.~(\ref{S-N}) as a
    function of $N$, for $Z=1$. Note that
    $S_{25}=2.4\times 10^{-6}$.}\label{fig3}}}
    \end{figure}

To confirm the consistency of our approximate calculation of
$\rh$, we check its Hermiticity. To do this we compare the
$N\times N$ matrices ${\cal Q}^{(N)}$ and ${\cal Q}^{(N)\dagger}$
defined in terms of their entries according to ${\cal
Q}^{(N)}_{mn}:=\br m|\rh|n\kt$ and ${\cal
Q}^{(N)\dagger}_{mn}:={{\cal Q}^{(N)}_{nm}}^*=\br n|\rh|m\kt^*$.
Clearly the condition $\rh^\dagger=\rh$ is equivalent to
    \be
    {\cal Q}^{(N)\dagger}\to{\cal Q}^{(N)}~~~~{\rm as}~~~~
    N\to\infty.
    \label{herm-condi}
    \ee
Noting that $\rH$ is not a bounded operator, we follow the method
of the preceding section and define the normalized matrix
$\hat{\cal Q}^{(N)}:= {\cal Q}^{(N)}/\parallel {\cal
Q}^{(N)}\parallel_2$. This allows us to identify the Hermiticity
condition~(\ref{herm-condi}) with
    \be
    S_N:=\parallel \hat{\cal Q}^{(N)}-\hat{\cal Q}^{(N)\dagger}
    \parallel_2\; \leq\, \nu_N.
    \label{S-N}
    \ee
Fig.~\ref{fig3} shows the plot of $S_N$ for $N\leq 25$ and $Z=1$.
Table~\ref{tab1} shows some typical values of $S_N$. The results
depicted in Fig.~\ref{fig3} and Table~\ref{tab1} are in complete
agreement with (\ref{S-N}).

Having obtained the matrix elements of $\delta \rH$ in the basis
$\{|n\kt\}$, we can compute its integral kernel,
    \be
    {\cal K}_{_{\delta \rH}}(\rx,\rx'):=\br\rx|\delta \rH|\rx'\kt,
    \label{kernel-delta-H}
    \ee
and the corresponding pseudo-differential operator in the
$\rx$-representation. In view of (\ref{z=0}), (\ref{n}) and
(\ref{delta-H-2}),
    \be
    {\cal K}_{_{\delta \rH}}(\rx,\rx')=
    \sum_{m,n=1}^N \delta \rH_{mn}^{(0)}\psi_m^{(0)}(\rx)
    \psi_n^{(0)}(\rx')^*=\sum_{m,n=1}^N \Delta_{mn}
    \sin\left[\frac{\pi m}{2}(\rx+1)\right]
    \sin\left[\frac{\pi n}{2}(\rx'+1)\right],
    \label{kernel-delta-H=}
    \ee
where
    \be
    \Delta_{mn}:= i^{\mu_m-\mu_n}\delta \rH_{mn}^{(0)}.
    \label{Delta}
    \ee

Next, we follow the derivation of Eqs.~(\ref{kernel-2}) --
(\ref{kernel-3}) to express $\delta \rH$ as a series in powers of
the momentum operator $\rp$. This yields
    \be
    \delta \rH=\sum_{\ell=0}^\infty \delta_\ell(\rx)\,\rp^\ell,
    \label{delta-H=2}
    \ee
where
    \bea
    \delta_\ell(\rx)&:=& i^\ell\, \tilde\delta_\ell(\rx),
    \label{delta-L}\\
    \tilde\delta_\ell(\rx)&:=&\frac{1}{\ell !}\int_{-1}^1
    {\cal K}_{_{\delta \rH}}(\rx,\rx')(\rx'-\rx)^\ell d\rx'=
    \sum_{m,n=1}^N \Delta_{mn}
    \sin\left[\frac{\pi m}{2}(\rx+1)\right] {\cal P}_{n\ell}(\rx),
    \label{tilde-delta-L}\\
    {\cal P}_{n\ell}(\rx)&:=&\frac{1}{\ell !}\int_{-1}^1
    \sin\left[\frac{\pi n}{2}(\rx'+1)\right](\rx'-\rx)^\ell d\rx'.
    \label{cal-P=}
    \eea
Using (\ref{sw}) -- (\ref{pwsw}) and (\ref{h=H+H}) we then obtain
    \be
    \rh\approxN
    \rH+\sum_{\ell=0}^\infty \delta_\ell(\rx)\,\rp^\ell=
    \rp^2+{\rm v}(\rx)+\sum_{\ell=0}^\infty
    \delta_\ell(\rx)\,\rp^\ell.
    \label{rh-sw=}
    \ee
The analogous expression for the Hermitian Hamiltonian $h$ associated
with the unscaled Hamiltonian $H$ is
    \be
    h\approxN\frac{p^2}{2m}+v(x)+\sum_{\ell=0}^\infty
    \gamma_\ell(x)\,p^\ell,
    \label{h-sw=}
    \ee
where
    \be
    \gamma_\ell(x):=\frac{L^{\ell-2}\delta_\ell(2x/L)}{m 2^{\ell-1}
    \hbar^{\ell-2}}.
    \label{gamma=}
    \ee

The integral in (\ref{cal-P=}) may be evaluated analytically. A
simple change of variable reduces it to an integral of the form
$\int_0^\pi y^m\sin(ny)dy$ that may be looked up in
\cite{Gradshteyn-Ruzhik}. Substituting the value of this integral
in (\ref{cal-P=}) and doing the necessary algebra, we find
    \be
    {\cal P}_{n\ell}(\rx)=\sum_{j=0}^\ell a_{nj\ell}(\rx+1)^{\ell-j},
    \label{cal-P=2}
    \ee
where
    \be
    a_{nj\ell}:=\frac{(-1)^\ell 2^{j+1}}{(\ell-j)!}
    \left\{\sum_{k=0}^{\lfloor\frac{j}{2}\rfloor}
    \frac{(-1)^{j+n+k+1}}{(\pi n)^{2k+1}(j-2k)!}+
    \frac{(-1)^{\lfloor\frac{j}{2}\rfloor}
    [1+(-1)^j]}{2(\pi n)^{j+1}}\right\},
    \label{coeff}
    \ee
and $\lfloor\frac{j}{2}\rfloor$ denotes the integer part of
$\frac{j}{2}$. As seen from (\ref{cal-P=2}), ${\cal P}_{n\ell}$ is
a polynomial of degree $\ell$.

Using (\ref{cal-P=2}) we can obtain a more explicit expression for
the coefficient functions $\delta_\ell$ appearing in
(\ref{rh-sw=}) and (\ref{gamma=}). Substituting (\ref{cal-P=2}) in
(\ref{tilde-delta-L}) and introducing
    \be
    b_{mk\ell}:=\sum_{n=1}^N\Delta_{mn}a_{nk\ell},
    \label{b-nkL}
    \ee
we have
    \be
    \tilde\delta_\ell(\rx)=\sum_{k=0}^\ell
    \sum_{m=1}^N b_{mk\ell}\;\sin\left[\frac{\pi m}{2}(\rx+1)\right]
    (\rx+1)^{\ell-k}.
    \label{delta=4}
    \ee
This relation together with (\ref{delta-L}) yield the desired
expression for $\delta_\ell(\rx)$.

\begin{figure}[tp]
    \centerline{\epsffile{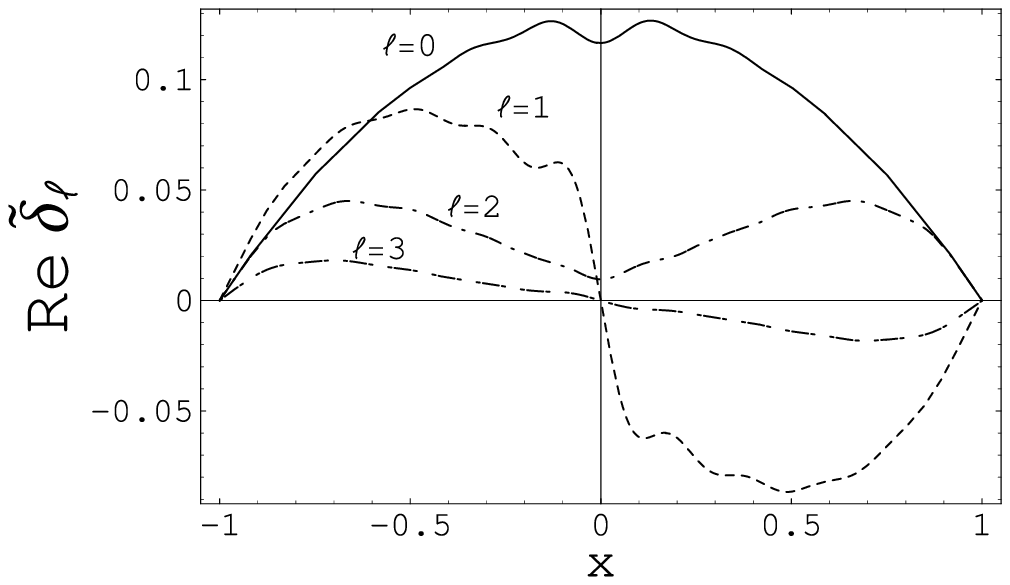}}
    \centerline{
    \parbox{12cm}{
    \caption{Plot of $\Re[\tilde \delta_\ell]$ for
    $\ell=0,1,2,3$, $Z=1$ and $N=20$.}\label{fig4}}}
    \vspace{2cm}
    \centerline{\epsffile{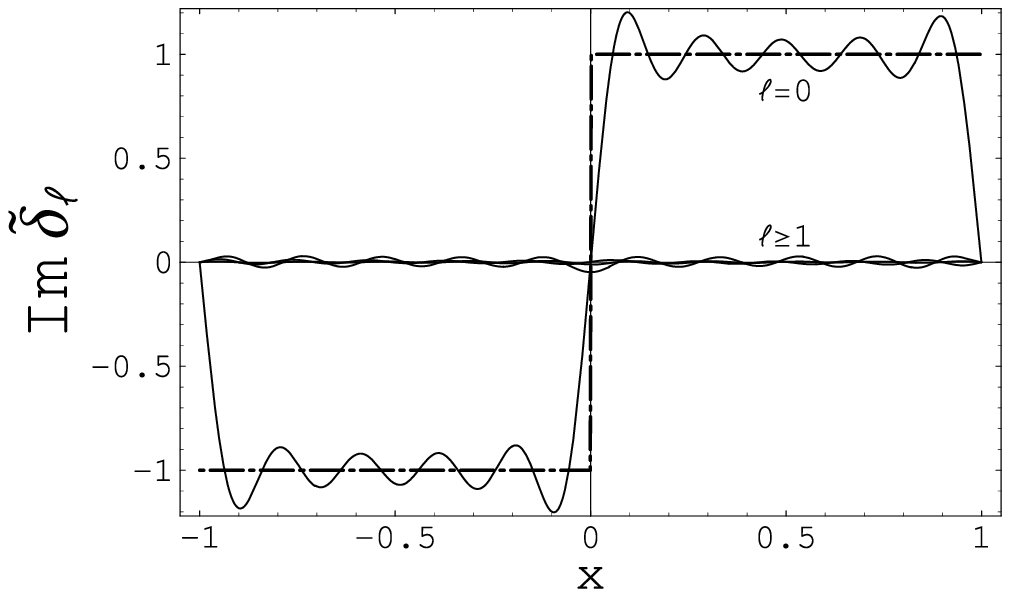}}
    \centerline{
    \parbox{12cm}{
    \caption{Plot of $\Im[\tilde \delta_\ell]$ for
    $\ell=0,1,2,3$, $Z=1$ and $N=20$. The dashed curve
    is the graph of $i{\rm v}$. The resemblance of the graphs of
    $\Im[\tilde \delta_0]$ and $i{\rm v}$ is described in the
    Appendix. The difference between $\Im[\tilde \delta_\ell]$ with
    $\ell\geq 1$ are too small to be distinguished in the
    energy scale determined by the potential which is unity in
    the units used.}\label{fig5}}}
    \end{figure}

Next, we recall that the standard calculation (\ref{exp-val}) of
the energy expectation value for a state vector $\psi\in{\cal
H}_{\rm phys}$, which uses the position wave function $\Psi$
introduced in Sec.~4, involves the representation $\hat \rh$ of
the Hermitian Hamiltonian $\rh$:
    \[\br\psi,H\psi\kt_+= \frac{2\hbar^2}{m L^2}\:
    \br\psi,\rH\psi\kt_+
    =\frac{2}{m L^2}\:\int_{-1}^1\Psi(\rx)^*\:\hat
    \rh\Psi(\rx)\,d\rx.\]
Using $\br \rx|\rh=\hat\rh\br\rx|$, the identity
    \be
    \br\rx|\rp=-i\frac{d}{d\rx}\br\rx|,
    \label{identity}
    \ee
and Eqs.~(\ref{delta-L}), (\ref{rh-sw=}), (\ref{delta=4}) and
(\ref{b-nkL}), we have
    \be
    \hat\rh\approxN -\frac{d^2}{d\rx^2}+{\rm v}(\rx)+
    \sum_{\ell=0}^\infty \tilde\delta_\ell(\rx)\;
    \frac{d^\ell}{d\rx^\ell}.
    \label{hat-h-sw}
    \ee
Figs.~\ref{fig4} and \ref{fig5} show the plots of the real and
imaginary parts of $\tilde \delta_\ell(\rx)$ for $\ell=0,1,2,3$,
$Z=1$ and $N=20$. As seen from Fig.~\ref{fig5}, the graph of
$\Im(\tilde\delta_0)$ is reminiscent of the approximation of a
step function, namely $i{\rm v}$, with the first few terms in its
Fourier series expansion. In the appendix we offer an explanation
for this observation. Furthermore, these figures suggest that
$\delta_\ell(\rx)$ and consequently $\tilde \delta_\ell(\rx)$ have
a vanishing large-$\ell$ limit. This can be established
analytically. Using (\ref{delta-L}) -- (\ref{cal-P=}) and the fact
that for all $\rx,\rx'\in[-1,1]$ both $(\rx-\rx')/2$ and $\sin[\pi
m(\rx'+1)/2]$ are bounded by $1$, we can easily conclude that
    \be
    |\tilde\delta_\ell(\rx)|\leq \frac{M 2^\ell}{\ell !},
    \label{bound-delta}
    \ee
where $M$ is a positive number depending on $\Delta_{mn}$. This
shows that, for all $\rx\in[-1,1]$, $\lim_{\ell\to\infty}
\tilde\delta_\ell(\rx)=0$.

\subsection{The Classical Hamiltonian}

Having obtained the Hermitian Hamiltonian $h$ for the
$PT$-symmetric square well, we can use the prescription described
in Sec.~3 to obtain the following expression for an underlying
classical Hamiltonian.
    \be
    H_c(x_c,p_c)\approxN \frac{p_c^2}{2m}+
    \sum_{\ell=0}^\infty \Re[\gamma_\ell(x_c)]\,p_c^\ell,
    \label{classs-H-sw}
    \ee
where $x_c\in[-L/2,L/2]$ and $p_c\in\R$. (\ref{classs-H-sw}) is a
meaningful relation, only if the series on its right-hand side
converges. In view of (\ref{delta-L}) and (\ref{gamma=}), the
latter is equivalent to the convergence of
    \be
    \sum_{\ell-0}^\infty \Re[\delta_\ell(\rx_c)]\rp_c^\ell,
    \label{delta-H-class}
    \ee
where $\rx_c\in[-1,1]$ and $\rp_c\in\R$. According to
(\ref{bound-delta}), for all $\rx_c\in[-1,1]$,
$|\Re[\delta_\ell(\rx_c)]|\leq M 2^\ell/\ell !$. Using this
relation (and performing the comparison and ratio tests
\cite{calculus}) we can easily show that the series
(\ref{delta-H-class}) converges (absolutely) for all values of
$\rp_c\in\R$ and $\rx_c\in[-1,1]$. Hence (\ref{classs-H-sw}) is a
meaningful expression yielding a well-defined classical
Hamiltonian.

Notice that $H_c$ as given by (\ref{classs-H-sw}) is not the
classical Hamiltonian in the strict sense that it would not
involve $\hbar$. This is simply because we have not evaluated the
$\hbar\to 0$ limit. Indeed, for the $PT$-symmetric square well
Hamiltonian~(\ref{sw}), the assumption that this limit exists has
drastic implications. This is simply because, according to
(\ref{rescale}), if we assume that the coupling constant $\zeta$
appearing in (\ref{pwsw-1}) does not depend on $\hbar$, then
taking the limit $\hbar\to 0$ corresponds to $Z\to\infty$. This
implies the occurrence of an infinite number of complex
eigenvalues which in turn indicates that the system does not admit
a unitary quantum mechanical description \cite{jmp-2004}. The only
way in which one can retain such a description and at the same
time be allowed to take the limit $\hbar\to 0$ is to assume that
$\zeta$ depends on $\hbar$ and is at least of order $\hbar^2$.

We can reach the same conclusion by noting that the condition
$Z<Z_\star\approx 4.48$, for the possibility of formulating a
unitary quantum theory for $PT$-symmetric square well, is
equivalent to $\zeta< 8.96 \hbar^2/(mL^2)< 2E_1 < E_2$, where
$E_1$ and $E_2$ are respectively the ground state and first
excited states of the system. For a light molecule, say O$_2$,
with mass $m\approx 30$~GeV, confined in a micron size well --
which should allow for a classical description -- we find
$\zeta<1.3\times 10^{-28}$~eV. This corresponds to the classical
molecule moving with a speed $v<4.7\times 10^{-20}$ m/s and a
temperature of $T<1.6\times 10^{-24}$~K. These numbers provide a
conclusive evidence that non-Hermiticity effects quantified with
the coupling constant $\zeta$ are quantum mechanical in nature and
have no classical counterpart.

Furthermore, recall that the classical limit $\hbar\to 0$ is
meaningful if it is accompanied with taking $n\to\infty$ in such a
way that $\hbar n$ stays constant.\footnote{This follows from
(\ref{en=}) and the requirement that in the classical limit not
all the energy levels collapse to zero.} But as we explained in
subsection~5.1, in the limit $n\to\infty$ the effects of the
non-Hermiticity of the Hamiltonian disappear. Therefore, the
classical limit of all the theories with different allowed values
of $Z<Z_\star$ coincides with that of the Hermitian infinite
square well $(Z=0)$.

The above discussion of the classical limit of the $PT$-symmetric
square well Hamiltonian is based on the requirement that the
corresponding quantum theory has a $\hbar\to 0$ limit. This is the
conventional way of defining the classical limit of a quantum
system. Yet we can consider the $\hbar$-dependent classical
observable $H_c$ and view it as a classical Hamiltonian with the
property that its pseudo-Hermitian quantization with appropriate
(symmetric) factor ordering yields the Hermitian Hamiltonian $h$.

There is also another approach for determining a classical
Hamiltonian for $PT$-symmetric quantum systems
\cite{bender-99,other-classical}. It involves a direct replacement
of the operators $x$ and $p$, that appear in the expression for
the quantum Hamiltonian operator $H$, by the classical position
$x_c$ and momentum $p_c$ and letting the latter take complex
values. If one applies this prescription to the $PT$-symmetric
square well and enforces the condition of the existence of a
proper $\hbar\to 0$ limit, then again the condition $Z<Z_\star$
implies $\zeta\to 0$, and one recovers the classical Hamiltonian
for a free particle confined in an infinite (real) square well.
However, if one does not identity the substitution $p\to p_c$ and
$x\to x_c$ with taking $\hbar\to 0$ in (\ref{sw}), then one
obtains a complex-valued `classical Hamiltonian', namely
    \be
    H'_c=\frac{p_c^2}{2m}+v(x_c).
    \label{H-prime}
    \ee
It is the classical Hamiltonian dynamical systems defined by such
complex `classical Hamiltonians' that are studied in
\cite{bender-99,other-classical}. Although we acknowledge the
interesting mathematical consequences of this study and its
relevance to the use of complex WKB approximation in calculating
the energy levels of various $PT$-symmetric models \cite{wkb}, we
are inclined to adopt the standard definition of a classical
observable which requires the latter to be real-valued
\cite{isham}.\footnote{This is because we are not aware of any
other precise definition of a classical observable.} The
$PT$-symmetric quantum mechanics also makes an implicit use of
this definition in insisting that the eigenvalues of the
observables, in particular the Hamiltonian, be real
\cite{bbj,cjp-2004b}. According to this definition, the
observables $x_c$ and $p_c$ assume real values, and $H'_c$, which
is a complex-valued function of $x_c$ and $p_c$, is not a physical
observable. In particular, it cannot serve as a physical classical
Hamiltonian (for a system with a one-dimensional configuration
space).

\subsection{Construction of the Observables}

The construction of the observables $O:{\cal H}_{\rm phys}\to{\cal
H}_{\rm phys}$ for the $PT$-symmetric square well mimics that of
the Hermitian Hamiltonian $h$. We begin our calculation of $O$ by
employing our approximation scheme to express (\ref{O=2}) in the
form
    \bea
    O&\approxN& \sum_{m,n=1}^N \sqrt{\frac{\epsilon_n}{\epsilon_m}}\,
    o_{mn}|\epsilon_m\kt\br\epsilon_n|+
    \sum_{m,n=N+1}^\infty o_{mn}^{(0)}|m\kt\br n|+\nn\\
    &&\sum_{m=1}^N\sum_{n=N+1}^\infty
    \left[{\frac{1}{\sqrt\epsilon_m}}
     o_{mn}|\epsilon_m\kt\br n|+
    \sqrt\epsilon_m\, o_{nm}
    |n\kt\br\epsilon_m|\right]\approxN o+\delta o,
    \label{O=5}
    \eea
where $o:{\cal H}\to{\cal H}$ is a Hermitian operator,
    \bea
    o_{mn}&:=&\br\epsilon_m|o|\epsilon_n\kt
    \approxN \left\{\begin{array}{ccc}
    \sum_{j,k=1}^N
    {\cal U}_{mj}^\dagger o_{jk}^{(0)}{\cal U}_{kn}&~~{\rm for}~~&
    m,n\leq N,\\
    \br\epsilon_m|o|n\kt\approxN \sum_{k=1}^N {\cal U}^\dagger_{mk}
    o^{(0)}_{kn}&~~{\rm for}~~&
    m\leq N,~n>N,\end{array}\right.
    \label{o-mn=}\\
    o^{(0)}_{mn}&:=&\br m|o|n\kt,
    \label{o-zero}\\
    \delta o&:=&\sum_{m,n=1}^N A_{mn}\,|m\kt\br n|+
    \sum_{m=1}^N\sum_{n=N+1}^\infty \left(B_{mn}\,|m\kt\br n|+
    C_{nm}\, |n\kt\br m|\right),
    \label{delta-o}\\
    A_{mn}&:=&\sum_{j,k=1}^N {\cal E}^{-1/2}_{mj}
    \,o_{jk}^{(0)}\,{\cal E}^{1/2}_{kn} -o^{(0)}_{mn},
    \label{coeff-A}\\
    B_{mn}&:=&\sum_{k=1}^N ({\cal E}_{mk}^{-1/2}-
    \delta_{mk})\,o^{(0)}_{kn},
    \label{coeff-B}\\
    C_{nm}&:=&\sum_{k=1}^N
    o^{(0)}_{nk}\,({\cal E}_{km}^{1/2}- \delta_{km}),
    \eea
$\delta_{mk}$ stands for the Kronecker delta function, and we have
used (\ref{Nth-order2}), (\ref{epsilon=}), and (\ref{root}).

We can express $\delta o$ and consequently $O $ as power series in
$\rp$ with $\rx$-dependent coefficients similarly to our
derivation of (\ref{h-sw=}). Note however that in this case we
have to deal with the infinite sum appearing in (\ref{delta-o}).
The presence of this sum is a manifestation of the fact that
(unlike the Hamiltonian) a general observable will mix the state
vectors $|n\kt$ with $n\leq N$ with those with $n>N$.

The power series expansion of $O$ in powers of $\rp$ has the form:
    \bea
    O&\approxN&o+\sum_{\ell=0}^\infty \omega_\ell(\rx)\: \rp^\ell,
    ~~~~~~~~\omega_\ell(\rx)=i^\ell\,\tilde\omega_\ell(\rx),
    \label{O=8}\\
    \tilde\omega_\ell(\rx)&:=&\sum_{k=0}^\ell
    \sum_{m=1}^\infty c_{mk\ell}\;\sin\left[\frac{\pi m}{2}(\rx+1)\right]
    (\rx+1)^{\ell-k},
    \label{omega=4}\\
    c_{mk\ell}&:=&\left\{\begin{array}{ccc}
    \sum_{n=1}^N i^{\mu_m-\mu_n}A_{mn}a_{nk\ell}+
    \sum_{n=N+1}^\infty i^{\mu_m-\mu_n}B_{mn}a_{nk\ell}&{\rm for}&
    m\leq N\\
    \sum_{n=1}^N i^{\mu_m-\mu_n} C_{mn}a_{nk\ell}&{\rm for}&
    m>N,\end{array}\right.
    \label{c-nkL}\\
    \eea
where $a_{nk\ell}$ are the coefficients given in (\ref{coeff}).
Again in calculating the expectation value of $O$ for a normalized
state vector $\psi\in{\cal H}_{\rm phys}$ with position wave
function $\Psi$, we use the position representation $\hat o$ of
$o$ (defined by $\br\rx|o=\hat o\br \rx|$):
    \be
    \br\psi,O\psi\kt_+=\int_{-1}^1\Psi(\rx)^*\:\hat o\,
    \Psi(\rx)\:d\rx.
    \label{position-exp-val}
    \ee
Note also that by replacing $(o,\rx,\rp)$ in (\ref{O=8}) with the
corresponding classical quantities $(o_c,\rx_c,\rp_c)$, we obtain
a generally complex-valued function $\Omega_c(\rx_c,\rp_c)$
(provided that the corresponding infinite series appearing in
(\ref{omega=4}) and (\ref{c-nkL}) converge.) Clearly, taking
$o=h$, we have $\Omega_c=H'_c$, where $H'_c$ is the complex
Hamiltonian~(\ref{H-prime}).

In order to compare the operators $o$ and $O$ we represent them in
the ordinary position representation, i.e., compare $\hat o$ with
$\hat O$. The latter is defined by $\br\rx|O=\hat O\br \rx|$.
Using (\ref{omega=4}), (\ref{c-nkL}), and (\ref{identity}), we
have
    \be
    \hat O=\hat o+\sum_{\ell=0}^\infty \tilde\omega_\ell(\rx)
    \:\frac{d^\ell}{d\rx^\ell}.
    \label{O=o+}
    \ee

\begin{figure}[ht]
    \centerline{\epsffile{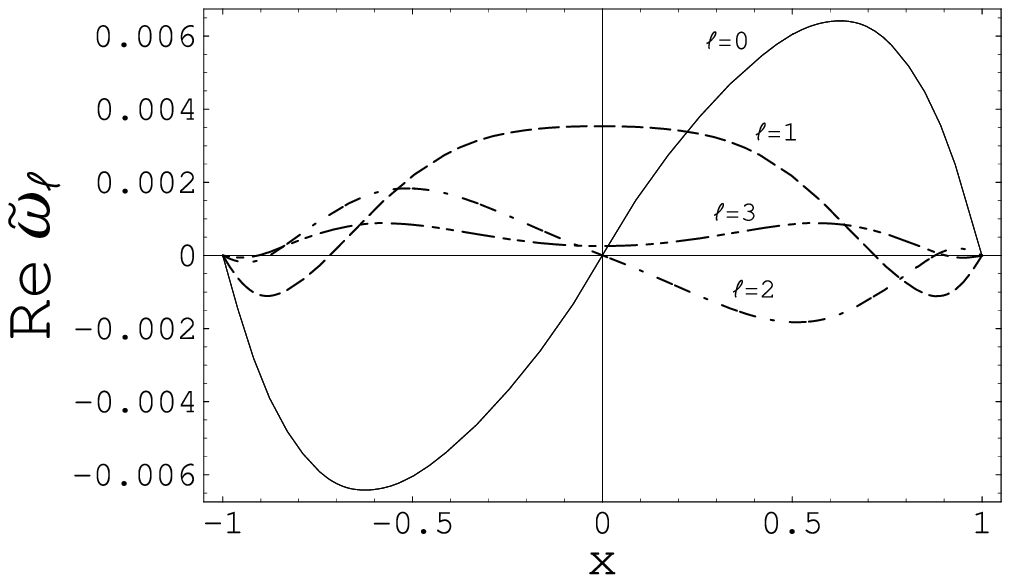}}
    \centerline{
    \parbox{12cm}{
    \caption{Graph of $\Re[\tilde\omega_\ell^{(X)}]$
    for $\ell=0,1,2,3$, $Z=1$, and $N=20$.}\label{fig6}}}
    \vspace{1cm}
    \centerline{\epsffile{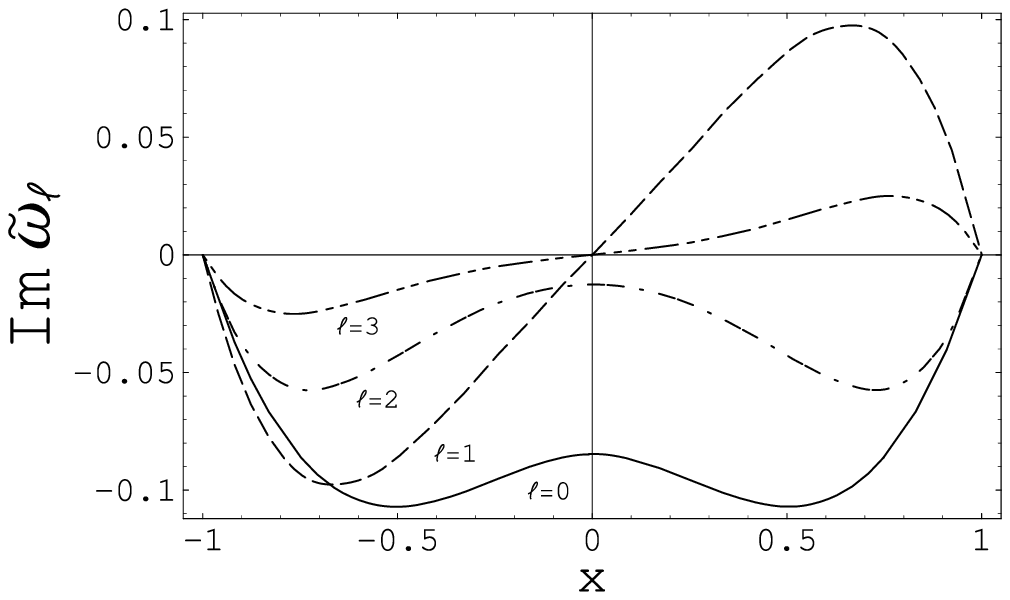}}
    \centerline{
    \parbox{12cm}{
    \caption{Graph of $\Im[\tilde\omega_\ell^{(X)}]$
    for $\ell=0,1,2,3$, $Z=1$, and $N=20$.}\label{fig7}}}
    \end{figure}

A concrete example is the dimensionless position operator ${\rm
X}:=2 X/L$: Letting $o=\rx$, we find
    \be
    {\rm X}=\rx+\sum_{\ell=0}^\infty \omega^{(X)}_\ell(\rx)
    \:\rp^\ell,
    \label{X=x+}
    \ee
where $\omega_\ell^{(X)}(\rx)$ denote the value of
$\omega_\ell(\rx)$ obtained by setting $o=\rx$ in (\ref{o-zero}).
The $\rx$-representation of ${\rm X}$ has the form
    \be
    \hat \rX=\rx+\sum_{\ell=0}^\infty \tilde\omega^{(X)}_\ell(\rx)
    \:\frac{d^\ell}{d\rx^\ell},
    \label{X=x+2}
    \ee
where $\tilde\omega^{(X)}_\ell(\rx)$ is the value of
$\tilde\omega_\ell(\rx)$ for $o=\rx$. Notice that the infinite
series in (\ref{c-nkL}) that defines $\tilde\omega_\ell(\rx)$
converges quite rapidly. This allows us to obtain an approximate
value for this series (for any value of $\ell$) by summing just
the first few terms. We can include enough terms in this series so
that the approximation error becomes smaller than our accuracy
index $\nu_N$. Figs.~\ref{fig6} and~\ref{fig7} illustrate the
plots of the real and imaginary parts of
$\tilde\omega_\ell^{(x)}$, for $\ell=0,1,2,3$, $Z=1$, and $N=20$,
that we have obtained in this way.

\subsection{Probability Density, Position Measurements, and
Localized States}

According to (\ref{prob-dens}), the probability density for the
localization in space is given by the modulus square of the
position wave function $\Psi$. We can employ our approximation
scheme to reduce the expansion (\ref{wf=epsilon}) of $\Psi$ into
the finite sum:
    \be
    \Psi(\rx)\approxN\psi(\rx)+\sum_{n=1}^N a_n
    \sin\left[\frac{\pi n}{2}(\rx+1)\right],
    \label{psi=5}
    \ee
where
    \bea
    a_n&=&i^{\mu_n}\left(\sum_{m=1}^N{\cal
    E}_{nm}^{1/2}f_m^{(0)}-f_n^{(0)}\right),
    \label{a-n=}\\
    f_m^{(0)}&=&\br m|\psi\kt=i^{-\mu_m}\int_{-1}^1
    \sin\left[\frac{\pi m}{2}(\rx'+1)\right]\psi(\rx')d\rx',
    \label{f-m-zero}
    \eea
and we have made use of (\ref{z=0}) -- (\ref{n}),
(\ref{Nth-order2}), (\ref{epsilon=}), (\ref{wf=epsilon}), and
(\ref{root}).

\begin{figure}[pt]
    \centerline{\epsffile{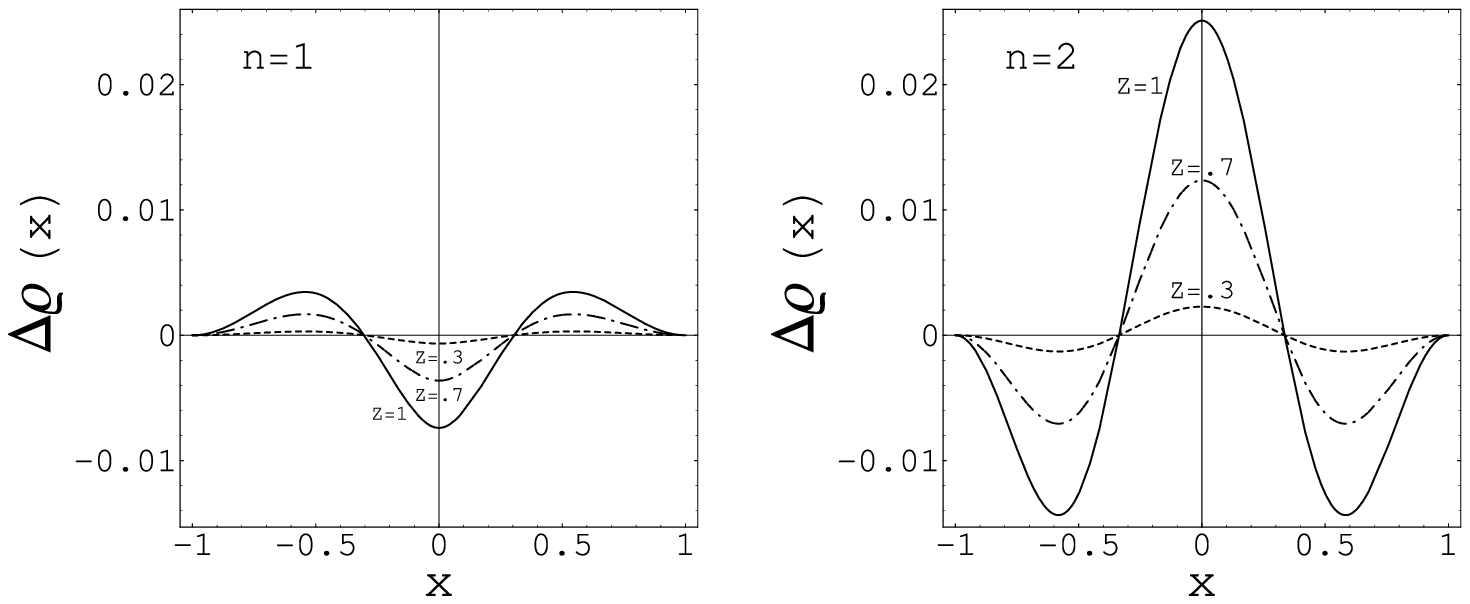}}
    \centerline{
    \parbox{15cm}{
    \caption{Graph of $\Delta\varrho=\varrho-\varrho_0$
    for $\psi(\rx)={\cal N}_n\sin[n\pi(\rx+1)/2]$ with $n=1,2$,
    $N=20$ and $Z=0.3,0.7,1$, where $\varrho$ is the probability
    density and $\varrho_0$ is its value for $Z=0$ and
    ${\cal N}_n$ are normalization constants.} \label{fig8}}}
    \vspace{1cm}
    \centerline{\epsffile{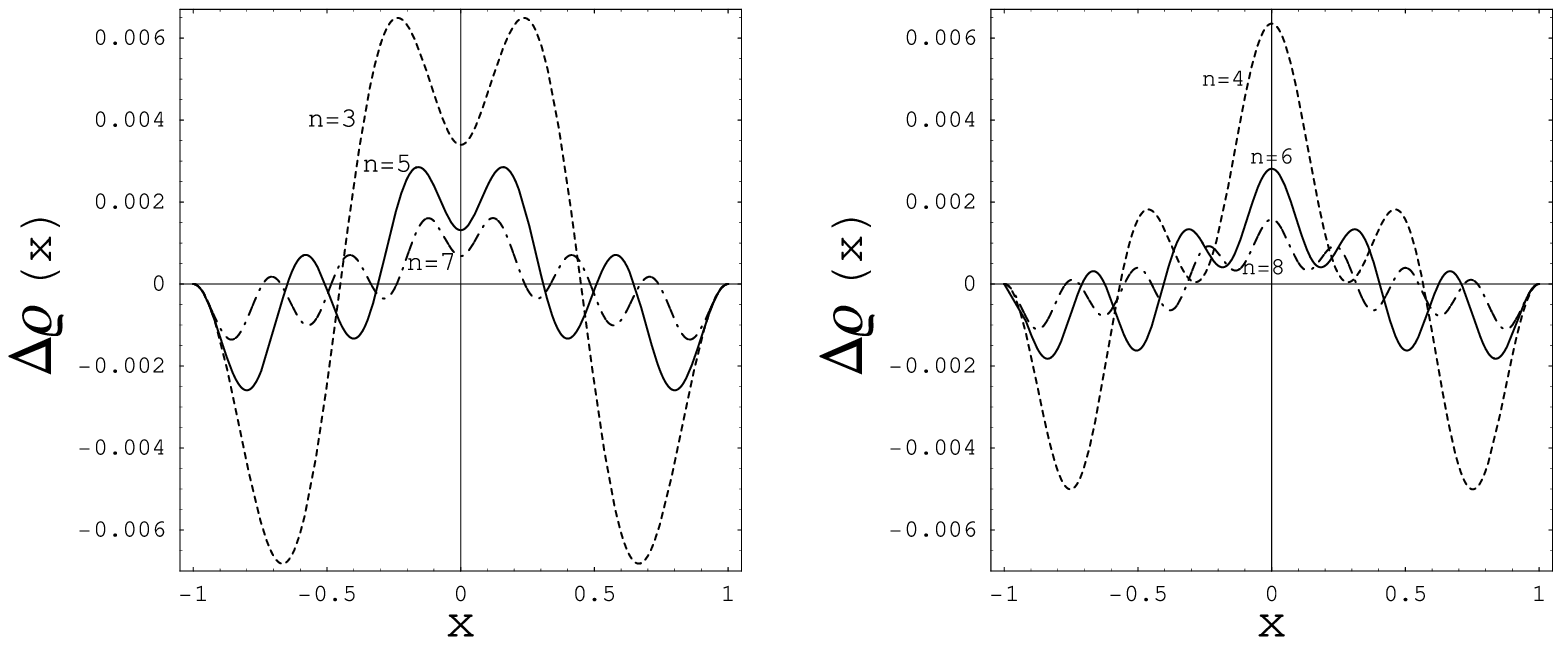}}
    \centerline{
    \parbox{15cm}{
    \caption{Graphs of $\Delta\varrho=\varrho-\varrho_0$ for
    $\psi(\rx)={\cal N}_n\sin[n\pi(\rx+1)/2]$ for
    $n=3,4,\cdots, 8$, $N=20$ and $Z=1$, where $\varrho$ is
    the probability density and $\varrho_0$ is its value for
    $Z=0$ and ${\cal N}_n$ are normalization constants.}\label{fig9}}}
    \end{figure}

Having obtained the general expression for the position wave
function, we can compute the probability density
$\varrho(\rx):=|\Psi(\rx)|^2$. Note, however, that the latter
expression is valid for the normalized wave functions.
Fig.~\ref{fig8} shows the plots of the difference $\Delta\varrho$
of the probability density $\varrho$ for $\psi(\rx)={\cal
N}_n\sin[n\pi(\rx+1)/2]$, with $n=1,2$, for $N=20$ and
$Z=0.3,0.7,1$ with that ($\varrho_0$) for $Z=0$. Fig.~\ref{fig9}
gives the plots of the probability density difference
$\Delta\varrho$ for $\psi(\rx)={\cal N}_n\sin[n\pi(\rx+1)/2]$ with
$n=3,4,\cdots,8$, $N=20$, and $Z=1$.\footnote{Here ${\cal N}_n$
are appropriate normalization constants.}

We can also use the expression (\ref{psi=5}) for the position wave
function to compute the position expectation value:
    \be
    \br\psi,{\rm X}\psi\kt_+=\int_{-1}^1 \rx |\Psi(\rx)|^2d\rx.
    \label{exp-valu-X}
    \ee
and the uncertainty in position
    \be
    \Delta \rx=\sqrt{\br\psi,{\rm X}^2\psi\kt_+-
    \br\psi,{\rm X}\psi\kt_+^2}.
    \ee
Table~\ref{tab2} gives $\Delta\rx$ for
$\psi(\rx)=N_n\sin[n\pi(\rx+1)/2]$ with $n=1,2,\cdots,7$,
$Z=0,0.5,1$ and $N=20$. It turns out the calculation of the same
quantities using $N=10$ yields results that differ from those
listed in Table~\ref{tab2} by numbers that are smaller than
$10^{-6}$. This is another confirmation of the consistency of our
approximation scheme. Furthermore, note that as we expect the
effect of the non-Hermiticity of the initial Hamiltonian
(\ref{sw}) diminishes as $n$ increases. Already for $n=6$, its
contribution to position uncertainty is smaller than the accuracy
index $\nu_{20}=0.0005$.
    \begin{table}[h]
    \begin{center}
  \begin{tabular}{||c||c|c|c|c||}
  \hline \hline
  $n$ & $Z=0$ & $Z=0.5$ & $Z=1$ &
  $\Delta\rx|_{Z=1}-\Delta\rx|_{Z=0}$\\
  \hline\hline
  1 & 0.3615 & 0.3618 & 0.3628 & $0.0013$ \\
  \hline
  2 & 0.5317 & 0.5308 & 0.5280 & $-0.0037$ \\
  \hline
  3 & 0.5575 & 0.5571 & 0.5559 & $-0.0016$ \\
  \hline
  4 & 0.5663 & 0.5660 & 0.5652 & $-0.0011$ \\
  \hline
  5 & 0.5704 & 0.5702 & 0.5697 & $-0.0006$ \\
  \hline
  6 & 0.5725 & 0.5724 & 0.5721 & $-0.0004$ \\
  \hline
  7 & 0.5738 & 0.5737 & 0.5735 & $-0.0003$\\
   \hline \hline\end{tabular}
  \end{center}
    \centerline{
    \parbox{14cm}{\caption{The position uncertainty $\Delta\rx$ for
    $\psi(\rx)=N_n\sin[n\pi(\rx+1)/2]$ with $n=1,2,\cdots,7$,
    $Z=0,0.5,1$, and $N=20$. Note that for $n=6$ and $7$ the
    difference between values of $\Delta\rx$ for $Z=1$ and
    $Z=0$ is smaller than the accuracy index $\nu_{20}=0.0005$.}
    \label{tab2}}}
    \end{table}

As seen from (\ref{X=x+}), in the $x$-representation the position
operator $X$ is a pseudo-differential operator. This in particular
means that the expectation value of $X$ in a state described by
the state vector $\psi$ depends on all the derivatives of $\psi$.
In this sense unlike the usual position operator, $X$ is a
nonlocal operator. Note however that this nonlocal character of
$X$ manifests itself only if one insists on using the usual
position representation $\psi(x)$ of the state vectors $\psi$.
This is not a reasonable choice, because being a non-Hermitian
operator acting in ${\cal H}_{\rm phys}$ the usual position
operator $x$ is not a physical observable.

Probably the best demonstration of the nonlocal nature of $X$ is
provided by the shape of the position state vector $\xi^{(y)}$
that is localized at $y\in(-L,L)$. As a function belonging to
${\cal H}$, it has the form
    \[\xi^{(y)}(x)=\br x|\xi^{(y)}\kt=\br x|\rho^{-1}|y\kt=
    \sum_{n=1}^\infty\epsilon_n^{-1/2}\varepsilon_n(x)
    \varepsilon_n(y)^*,\]
where we have employed (\ref{xi=}), (\ref{resol-2}), and
(\ref{epsilon-x}). Using the same method as the one leading to
(\ref{psi=5}) we can express this relation as
    \be
    \xi^{(y)}(x)\approxN\delta(x-y)+{\cal F}(x,y),
     \label{local=xy}
    \ee
where
    \[{\cal F}(x,y):=\sum_{m,n=1}^N
    i^{\mu_m-\mu_n}({\cal E}^{-1/2}_{mn}-\delta_{mn})
    \sin\left[\frac{\pi m}{L}(x+\frac{L}{2})\right]
    \sin\left[\frac{\pi n}{L}(y+\frac{L}{2})\right]
    ,\]
and we have used (\ref{root}).

\begin{figure}[hp]
    \centerline{\epsffile{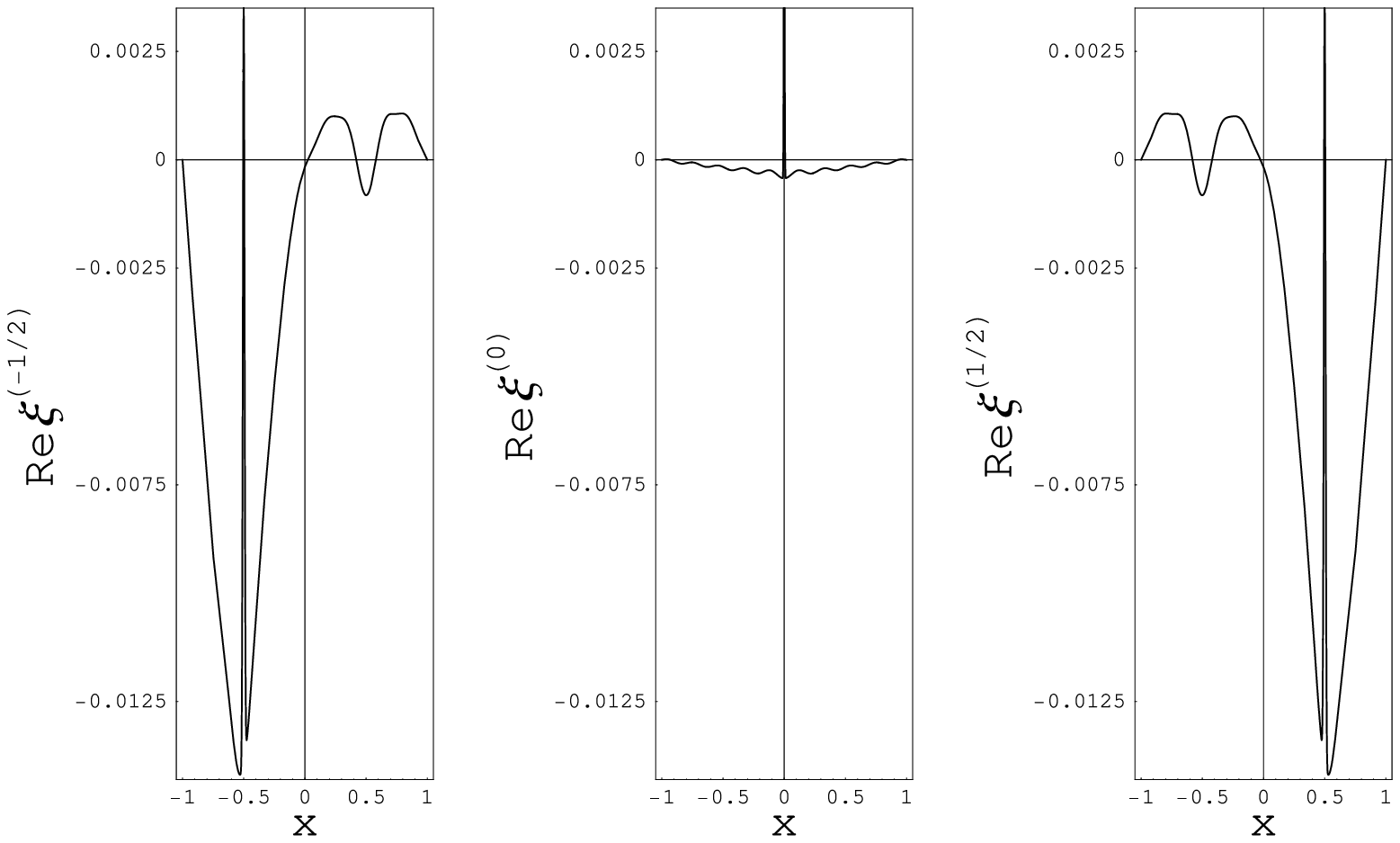}}
    \vspace{1cm}
    \centerline{\epsffile{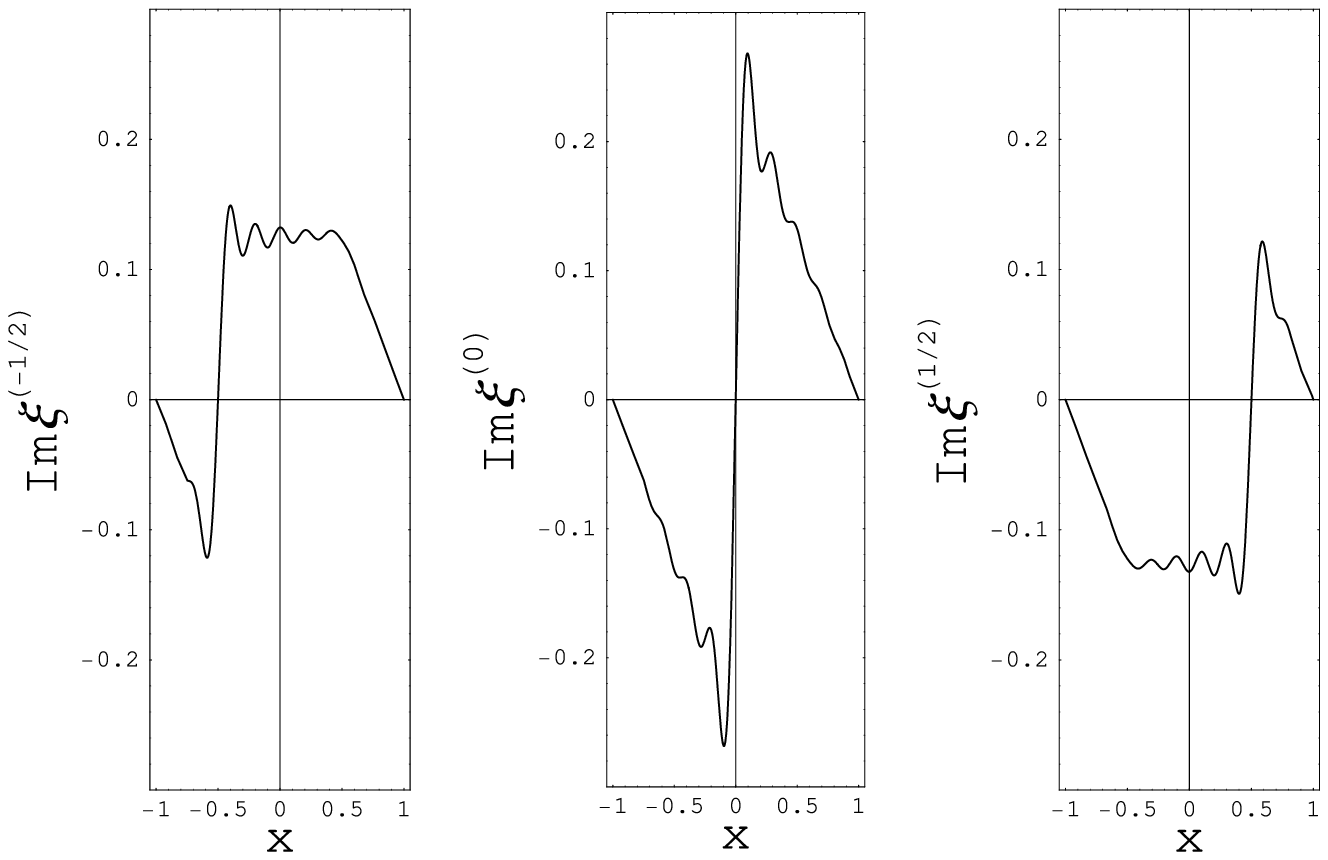}}
    \centerline{
    \parbox{15cm}{
    \caption{Graph of $\Re[\xi^{(y)}]$ and
    $\Im[\xi^{(y)}]$ for $L=2$, $N=20$, $Z=1$, and
    $y=-1/2,0,1/2$. Note that $\Re[\xi^{(y)}]$ has a
    $\delta$-function singularity at $y$ and that except for
    this singularity the scale of variations of
    $\Im[\xi^{(y)}]$ is much greater than that of
    $\Re[\xi^{(y)}]$.}\label{fig10}}}
    \vspace{1cm}
    \end{figure}

\begin{figure}[t]
    \centerline{\epsffile{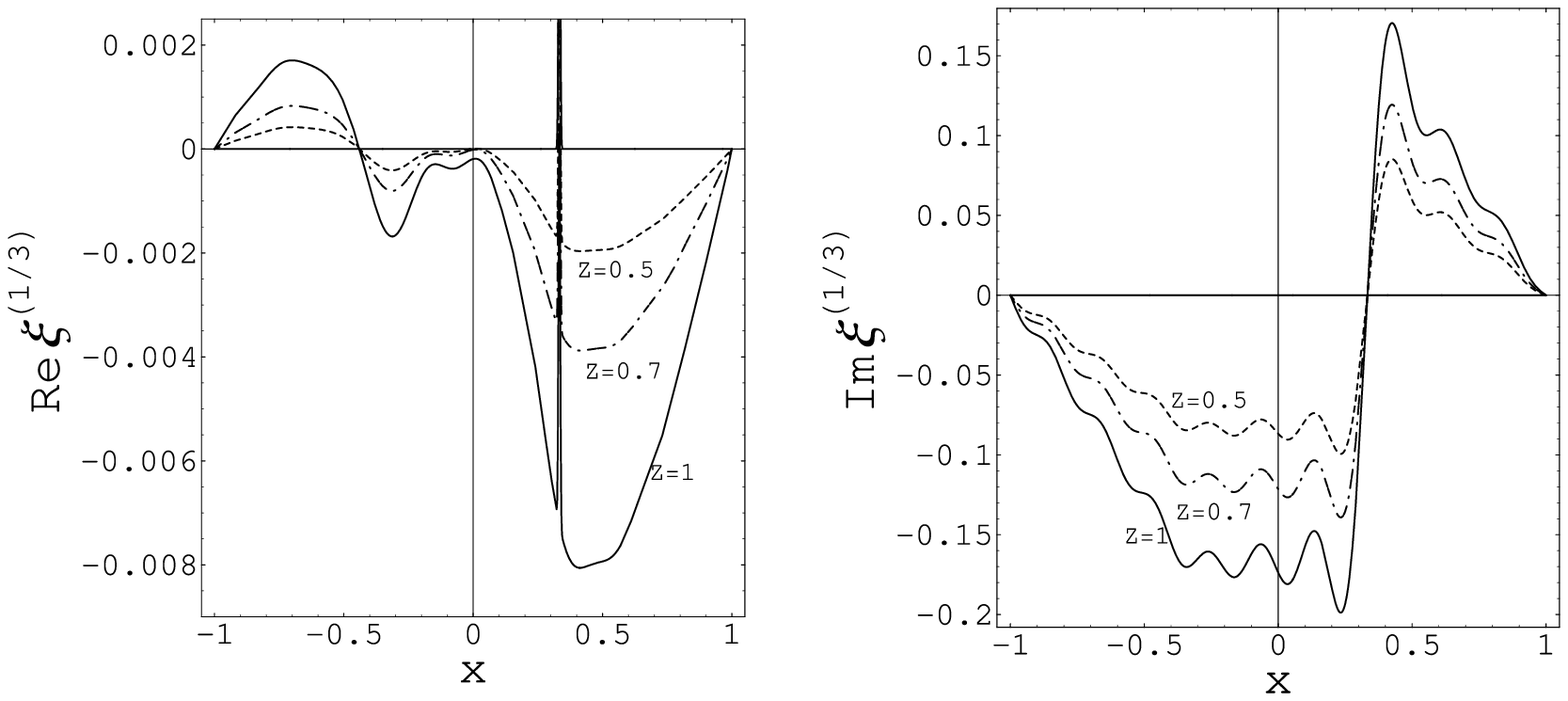}}
    \centerline{\parbox{15cm}{
    \caption{Graph of $\Re[\xi^{(1/3)}]$ and
    $\Im[\xi^{(1/3)}]$ for $L=2$, $N=20$, and
    $Z=0,0.5,0.7,1$.}\label{fig11}}}
    \vspace{1cm}
    \end{figure}

Fig.~\ref{fig10} shows the real and imaginary parts of
$\xi^{(y)}(x)$ for $L=2$, $N=20$, $Z=1$, and $y=-1/2,0,1/2$.
Fig.~\ref{fig11} shows the real and imaginary parts of
$\xi^{(1/3)}(x)$ for $L=2$, $N=20$, and various values of $Z$.
As expected the spreading of the localized state $\xi^{(1/3)}$ is
an increasing function of the non-Hermiticity parameter $Z$.

\subsection{Dynamical Consequences of Non-Hermiticity}

In the preceding subsection we discussed the computation of the
observables and the associated physical quantities. These provide
information on the kinematical content of the $PT$-symmetric
square well. In this subsection, we investigate its dynamical
content.

The time evolution of an initial state vector $\psi(t_0)\in{\cal
H}_{\rm phys}$ is given by
    \be
    \psi(t)=e^{-i(t-t_0)H/\hbar}\psi(t_0),~~~~~~\forall t\in\R.
    \label{time-evo}
    \ee
Alternatively, in terms of the dimensionless time parameter:
    \be
    \tau:=\left(\frac{2\hbar}{mL^2}\right)\,t,
    \label{tau=t}
    \ee
we have
    \be
    \psi(\tau)=e^{-i(\tau-\tau_0)\rH}\psi(\tau_0),
    \label{time-evo-tau}
    \ee
where $\tau_0=2\hbar t_0/(mL^2)$.

Expanding $\psi(\tau)$ in the basis $\{\psi_n\}$, we can express
(\ref{time-evo-tau}) in the form
    \be
    \psi(\tau)=\sum_{n=1}^\infty c_n
    e^{-i(\tau-\tau_0)\rE_n}
    \psi_n,~~~~~~~~~~c_n:=\br\phi_n|\psi(\tau_0)\kt.
    \label{psi-expand}
    \ee
Next, we employ our $N$-th order approximation scheme. Using
(\ref{Nth-order}), we have
    \be
    \psi(\tau)\approxN \sum_{n=1}^N c_n
    e^{-i(\tau-\tau_0)\rE_n}
    \psi_n+\sum_{n=N+1}^\infty c^{(0)}_n
    e^{-i\pi^2n^2(\tau-\tau_0)/4}\psi_n^{(0)},
    \label{psi-exand-approx}
    \ee
where
    \be
    c^{(0)}_n:=\br\psi_n^{(0)}|\psi(\tau_0)\kt,
    \label{c-n-zero}
    \ee
and we have used $\rE_n\approxN\rE_n^{(0)}:=\pi^2n^2/4$ for $n>N$.

To explore the dynamical effects of the non-Hermiticity of the
Hamiltonian (for $Z\neq 0$) we compute the position expectation
value for the evolving state vectors $\psi(\tau)$ having the
initial value:
    \be
    \psi(\tau_0)={\cal N}_j \psi_j^{(0)}~~~~~{\rm with}~~~~~j\leq N,
    \label{ini-psi}
    \ee
and ${\cal N}_j:=\br\psi_j^{(0)},\psi_j^{(0)}\kt_+^{-1/2}$. For
these choices of the initial state vector, $c^{(0)}_n=0$ for $n>N$
and (\ref{psi-exand-approx}) simplifies as
    \be
    \psi(\tau)\approxN \sum_{n=1}^N c_n
    e^{-i(\tau-\tau_0)\rE_n} \psi_n
    \label{simple}
    \ee
Furthermore, for $Z=0$, $\psi(\tau_0)$ corresponds to a stationary
state with a vanishing position expectation value for all
$\tau\in\R$. For $Z>0$, $\psi(\tau_0)$ does not represent a
stationary state and the position expectation value is a
nonconstant function of time. To determine this function, we use
the position representation of the state.

\begin{figure}[p]
    \centerline{\epsffile{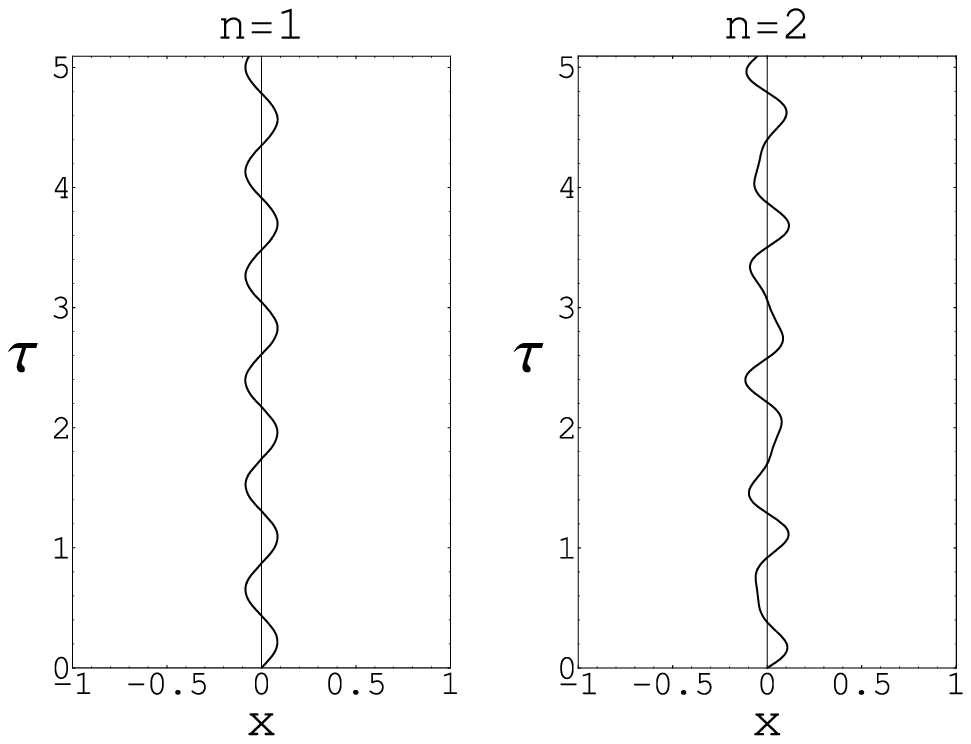}}
    \centerline{
    \parbox{16cm}{
    \caption{Graph of the trajectory traced by the
    position expectation value in time for the initial
    state vector $\psi(0)={\cal N}_j \psi_j^{(0)}$ with
    $j=1,2$, where $Z=1$, $N=10$, $\tau_0=0$, $\tau\in[0,16/\pi]$.
    The horizontal and vertical axes respectively represent
    $\br\psi(\tau),\rX\psi(\tau)\kt_+$ and $\tau$. The
    $\tau$-axis also corresponds to the trajectory for the
    Hermitian case ($Z=0$). Notice that $16/\pi\approx 5.1$ is twice the
    characteristic period $2\pi/\rE_1^{(0)}$ for the ground
    state of the corresponding Hermitian square well.}
    \label{fig12}}}
    \vspace{1cm}
    \centerline{\epsffile{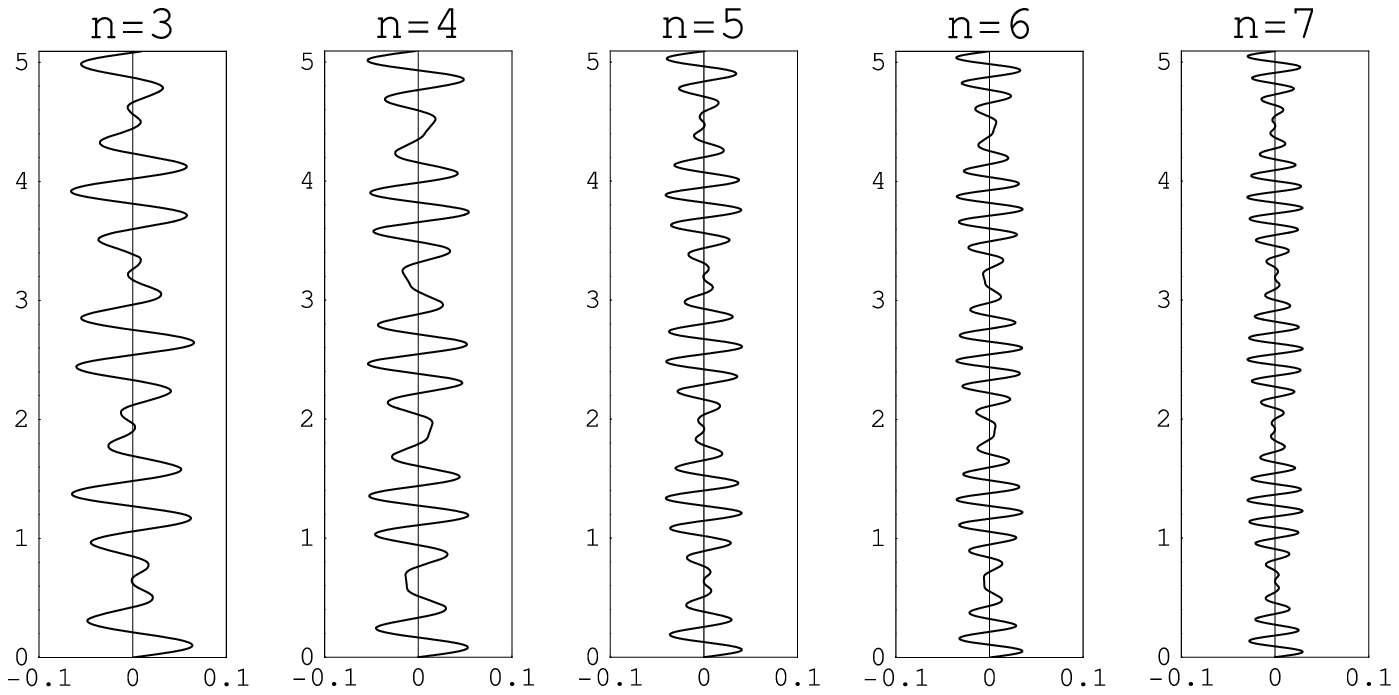}}
    \centerline{
    \parbox{16cm}{
    \caption{Graph of the trajectory traced by
    $\br\psi(\tau),\rX\psi(\tau)\kt_+$ for $\psi(0)=
    {\cal N}_j \psi_j^{(0)}$ with $j=3,4,\cdots,7$ and the same
    parameters and conventions as in Fig.~\ref{fig12}. Note that
    the range of values of the horizontal axis is reduced to
    amplify the behavior of the trajectories. The envelops
    seem to have the same period as that of $E_1^{(0)}$, i.e.,
    $2\pi/\rE_1^{(0)}\approx 8/\pi$.}
    \label{fig13}}}
    \end{figure}

Let $\Psi(\rx;\tau):=\br\xi^{(\rx)},\psi(\tau)\kt_+=
\br\rx|\rho|\psi(\tau)\kt$ be the position wave function for the
state vector $\psi(\tau)$. Then, in light of (\ref{Nth-order2}),
(\ref{epsilon=}), (\ref{root}), and (\ref{simple}), we have
    \be
    \Psi(\rx;\tau)\approxN \sum_{n=1}^N c_n\,
    e^{-i(\tau-\tau_0)\rE_n}\, \Gamma_n(\rx),
    \label{wf-tau}
    \ee
where for all $n\leq N$
    \be
    \Gamma_n(\rx):=\br\rx|\rho|\psi_n\kt
    \approxN \sum_{q,k=1}^N i^{\mu_q}
    {\cal E}_{qk}^{1/2}\,\br k|\psi_n\kt\,
    \sin\left[\frac{\pi q}{2}\,(\rx+1)\right].
    \label{Gamma-tau}
    \ee
Next, we employ (\ref{exp-valu-X}) to compute the position
expectation value for $\psi(\tau)$. In view of (\ref{wf-tau}),
    \be
    \br\psi(\tau),\rX\psi(\tau)\kt_+= \int_{-1}^1
    \rx|\Psi(\rx;\tau)|^2 d\rx\approxN \sum_{m,n=1}^N
    \Theta_{mn}\, e^{-i(\tau-\tau_0)(\rE_n-\rE_m)},
    \label{exp-val-tau}
    \ee
where
    \bea
    \Theta_{mn}&:=&c_m^*c_n\int_{-1}^1 \rx\, \Gamma_m(\rx)^*
    \Gamma_n(\rx)\,d\rx.
    \nn\\
    &=&c_m^*c_n\sum_{k,q,v=1}^N\sum_{\stackrel{u=1}{u\neq q}}^N
    \left(\frac{8 q u [(-1)^{q+u}-1]}{\pi^2
    (q^2-u^2)^2}\right)
    i^{\mu_u-\mu_q}
    {\cal E}_{kq}^{1/2}{\cal E}_{uv}^{1/2}\,
    \br\psi_m|k\kt\br v|\psi_n\kt.
    \label{theta-tau}
    \eea
Figs.~\ref{fig12} and \ref{fig13} show the trajectories traced by
the position expectation value (\ref{exp-val-tau}) in time, for
the initial state vector (\ref{ini-psi}) with $j=1,2,\cdots,7$,
$Z=1$ and $N=10$.

\textheight = 23.8cm

\section{Discussion and Conclusion}

In this article we have outlined a general formulation of
$PT$-symmetric (and more generally pseudo-Hermitian) quantum
mechanics paying attention to the physical aspects of the theory.
This formulation is consistent with the requirements of quantum
measurement theory and allows for the determination of the
physical observables. In fact, to the best of our knowledge, this
paper is the first to offer an explicit calculation of observables
and concrete physical quantities for a $PT$-symmetric system with
an infinite-dimensional Hilbert space. Perhaps more importantly,
it proposes a method to identify an underlying classical
Hamiltonian that satisfies the usual postulates of classical
mechanics and a quantization scheme that relates the latter to the
defining Hamiltonian of the theory. We view this as a necessary
step toward a clearer understanding of the potential physical
applications of $PT$-symmetric quantum mechanics.

Another important outcome of our investigation is that we are now
able to consider the addition of the interaction terms to a
$PT$-symmetric Hamiltonian $H$ without disturbing the structure of
its Hilbert space. This is simply done by selecting the additional
interaction terms from among physical observables.

Our investigation of the $PT$-symmetric square well revealed the
fact that the underlying classical Hamiltonian for this system
coincides with that of the ordinary Hermitian infinite square
well. In other words, the non-Hermiticity effects are quantum
mechanical in nature. This can be traced back to the simple
observation that the non-Hermiticity of the Hamiltonian~(\ref{sw})
only affects the low-lying energy levels.\footnote{We do not claim
that this is a common feature of all the $PT$-symmetric quantum
Hamiltonians. In general, we expect the non-Hermiticity of the
quantum Hamiltonian to affect the underlying classical
Hamiltonian. We leave a more detailed study of this issue for a
future publication.}

Our general results confirm the assertion that as a fundamental
theory $PT$-symmetric quantum mechanics is both mathematically and
physically equivalent to conventional quantum mechanics
\cite{critique}. In fact it is this very equivalence that allows
for the computation of the physical observables. This in turn
leads to the natural question whether there is any valid
motivation for further development of $PT$-symmetric and
pseudo-Hermitian quantum mechanics. Our answer to this question is
in the affirmative. It is supported by the following observations.
    \begin{enumerate}
    \item
As we showed for any $PT$-symmetric Hamiltonian $H$, there is a
corresponding Hermitian Hamiltonian. But the latter is a
generically nonlocal (pseudo-differential) operator. Therefore, if
one is interested in calculating physical quantities that make
explicit use of the Hamiltonian one is naturally inclined to make
use of the original Hamiltonian $H$ and the inner product
$\br\cdot,\cdot\kt_+$. However, if one wishes to compute
quantities involving the position and momentum of the system, then
one is essentially forced to use the Hermitian picture. In
summary, developing $PT$-symmetric quantum mechanics opens up the
possibility of treating quantum systems with certain nonlocal
Hermitian Hamiltonians. The classical Hamiltonian for such a
system is a real analytic function of $x$ and $p$ that involves
arbitrarily high powers of $p$. The pseudo-Hermitian quantization
scheme introduced in this paper provides a description of the
quantum systems associated with these complicated Hamiltonians. It
yields a $PT$-symmetric quantum Hamiltonian operator that is a
local (differential) operator. In a sense the use of the
$PT$-symmetric quantum mechanics is equivalent to trading a
complicated nonlocal Hermitian Hamiltonian with a local
$PT$-symmetric Hamiltonian.\footnote{The characterization of all
the nonlocal Hermitian Hamiltonians that may be mapped to
PT-symmetric or pseudo-Hermitian Hamiltonians of the standard
(kinetic+potential) form is a difficult open problem.}
    \item
The basic ideas so far developed within the framework of
pseudo-Hermitian quantum mechanics to assess the structure of
$PT$-symmetric quantum mechanics have some remarkable applications
in relativistic quantum mechanics \cite{cqg,rqm1,rqm}, quantum
cosmology \cite{cqg,ap}, statistical mechanics \cite{ahmed}, and
magnetohydrodynamics \cite{dynamo}. This strengthens the belief
that the study of $PT$-symmetric quantum mechanics may lead to
some concrete advances in other research areas. It is needless to
mention the possibility that the field theoretical extension of
such a study may actually turn out to achieve some of the
ambitious goals described in \cite{bbj-ajp}.
    \end{enumerate}

Next, we wish to elucidate the relationship between our approach
and the formulation of the $PT$-symmetric quantum mechanics based
on the so-called charge-conjugation operator $C$ as outlined in
\cite{bbj}. See also \cite{weigert}. In our approach the metric
operator $\eta_+$ plays the same role as the operator $C$. In fact
as shown in \cite{p9}, $C$ may be expressed in terms of $\eta_+$
according to
    \be
    C=\eta_+^{-1}P.
    \label{C=}
    \ee
Although both $\eta_+$ and $C$ determine the inner product of the
physical Hilbert space, the expression of the latter in terms of
$\eta_+$ is slightly simpler. It also agrees with the standard
mathematical approach used in dealing with different inner
products on the same vector space. Furthermore, the recent
attempts \cite{bender-2004} at approximate calculations of $C$ for
$PT$-symmetric potentials of the form $\mu^2x^2-\lambda^2 (ix)^N$,
with $\mu,\lambda\in\R$, has revealed the remarkable fact that
these calculations simplify enormously provided that one first
computes $\eta_+$ and then uses (\ref{C=}) to determine
$C$.\footnote{What authors of \cite{bender-2004} do is to express
$\eta_+$ as $e^{-Q}$, calculate $Q$ approximately, and express $C$
as $C=e^Q P$. Note that this equation is identical with
Eq.~(\ref{C=}) that was initially derived in \cite{p9}. The
calculation of $Q$ makes use of the fact that $C$ is a symmetry
generator. The observation that operators of the form
$\eta_1^{-1}\eta_2$ generate symmetries of a Hamiltonian that is
both $\eta_1$- and $\eta_2$-pseudo-Hermitian was initially
reported in \cite{p1}.} This provides a practical justification
for the assertion that $\eta_+$ is a more basic ingredient of the
theory than $C$. The construction of the observables provides a
much more concrete evidence for the validity of this assertion.
Note also that the formulation of the theory that uses $C$ and
avoids any explicit mention of $\eta_+$ leads to the same general
conclusions such as the physical equivalence of the $PT$-symmetric
and conventional quantum mechanics. A mathematically rigorous
proof of this statement is given in \cite{cjp-2004b}.

Finally, we wish to comment on whether one can apply the general
scheme offered by pseudo-Hermitian quantum mechanics to
$PT$-symmetric systems defined on a complex contour. The negative
attitude expressed by some of the workers on this issue is based
on the argument that for these systems the eigenfunctions of the
Hamiltonian do not belong to the Hilbert space $L^2(\R)$ and hence
one cannot define a metric operator $\eta_+$ and apply the results
of the theory of pseudo-Hermitian operators. The problem with this
argument is that nowhere in the formulation of the
pseudo-Hermitian quantum mechanics \cite{cjp-2003} does one assume
that the initial Hilbert space ${\cal H}$ is $L^2(\R)$. As
explained in Sec.~2, ${\cal H}$ is constructed in two steps:~(i)
One takes the span $V_H$ of the eigenvectors $\psi_n$ of $H$ (that
is assumed to have a real and discrete spectrum) and endows it
with some arbitrary (positive-definite) inner product. (ii)~One
performs the Cauchy completion of this inner product space to
obtain the Hilbert space ${\cal H}$. The only important condition
to be checked is whether ${\cal H}$ is separable. This follows
from the following simple argument (see also \cite{cjp-2004b}.)
The set $\{\psi_n\}$ of the eigenvectors spans $V_H$, and as
${\cal H}$ is the Cauchy completion of $V_H$, $V_H$ is dense in
${\cal H}$. This implies that ${\cal H}$ is the closure of the
span of $\{\psi_n\}$. Being eigenvectors with different
eigenvalues, $\psi_n$ are also linearly independent. Hence
$\{\psi_n\}$ is a countable basis of ${\cal H}$. In particular,
performing Gram-Schmidt orthonormalization on $\{\psi_n\}$, one
can construct a countable orthonormal basis of ${\cal H}$. This is
equivalent to the statement that ${\cal H}$ is a separable Hilbert
space \cite{reed-simon}. This general argument shows that indeed
there is no obstruction to employ pseudo-Hermitian quantum
mechanics to systems defined on a complex contour. The practical
difference with systems defined on the real axis is that one
cannot make a direct use of the familiar $L^2$-inner product. It
turns out that this does not lead to any insurmountable difficulty
either. On the contrary, the use of the machinery of
pseudo-Hermitian quantum mechanics in describing $PT$-symmetric
systems defined on a complex contour has both practical and
conceptual advantages \cite{p62}.

\section*{Acknowledgments}

This work has been supported by the Turkish Academy of Sciences in
the framework of the Young Researcher Award Program
(EA-T$\ddot{\rm U}$BA-GEB$\dot{\rm I}$P/2001-1-1). A.\ M.\ wishes
to thank Tekin Dereli and Ali \"Ulger for fruitful discussions.

\newpage

\begin{appendix}

\section{Appendix}

In this appendix we present a general method for checking the
Hermiticity of the Hamiltonian $\rh$ using its power series
expansion (\ref{rh-sw=}). This provides an interesting explanation
for the resemblance of the graph of the function $\Im[
\tilde\delta_0(\rx)]$ to that of the $i{\rm v}(\rx)$ as shown in
Fig.~\ref{fig5}.

Using (\ref{rh-sw=}) and the fact that ${\rm v}(\rx)$ is
imaginary, we have
    \be
    \rh^\dagger\approxN \rp^2-{\rm v}(\rx)+\sum_{\ell=0}^\infty
    \rp^\ell\,\delta_\ell(\rx)^*.
    \label{h-dagger}
    \ee
Substituting (\ref{rh-sw=}) and (\ref{h-dagger}), in
$\rh^\dagger=\rh$, we then find
    \be
    {\rm v}(\rx)+\frac{1}{2}\sum_{\ell=0}^\infty[\delta_\ell(\rx)
    \rp^\ell-\rp^\ell\,\delta_\ell(\rx)^*]\approxN 0.
    \label{hermiticity-h}
    \ee
Our purpose is to express the left-hand side of this relation as a
power series in $\rp$ with all $\rx$-dependent coefficient
appearing to the left of powers of $\rp$, i.e., obtain a set of
functions $f_\ell$ such that
    \be
    \sum_{\ell=0}^\infty f_\ell(\rx)\,\rp^\ell=
    {\rm v}(\rx)+\frac{1}{2}\sum_{\ell=0}^\infty[\delta_\ell(\rx)
    \rp^\ell-\rp^\ell\,\delta_\ell(\rx)^*]=
    {\rm v}(\rx)+\frac{1}{2}\sum_{\ell=0}^\infty
    i^\ell[\tilde\delta_\ell(\rx)
    \rp^\ell-\rp^\ell\,\tilde\delta_\ell(\rx)^*].
    \label{f=}
    \ee
Then the condition (\ref{hermiticity-h}) takes the form
    \be
    f_\ell(\rx)\approxN 0~~~~~~~{\rm for~all}~~\ell\in\{0,1,2,\cdots\}.
    \label{f=zero}
    \ee

In order to compute $f_\ell$, we use the following useful identity
which may be proven by induction on $\ell$ and use of
$[\rx,\rp]=i$.
    \be
    \rp^\ell f(\rx)=\sum_{k=0}^\ell \mbox{\scriptsize$
    \left(\begin{array}{c}
    \ell\\ k\end{array}\right)$}\, (-i)^k\,\frac{d^k
    f(\rx)}{d\rx^k}\, \rp^{\ell-k},
    \label{commutator-id}
    \ee
where $f:\R\to\R$ is an $\ell$-times differentiable function and
{\scriptsize $\left(\begin{array}{c} \ell\\
k\end{array}\right)$}$:= \frac{\ell !}{k!(\ell-k)!}$.

Using~(\ref{commutator-id}), (\ref{f=}), and the fact that ${\rm
v}(\rx)$ is imaginary, we can express the condition
(\ref{f=zero}), after some rather lengthy algebra, as follows.
\begin{itemize}
\item For $\ell=0$, $f_0\approxN 0$ yields:
    \be
    u_0(\rx)\approxN 0,~~~~~~~~w_0(\rx)\approxN i{\rm v}(\rx),
    \label{condi-zero}
    \ee
where
    \bea
    u_0(\rx)&:=&\sum_{k=1}^{\infty} (-1)^{k}\frac{d^k}{d\rx^k}\,
    \Re[\tilde\delta_k(\rx)],
    \label{u-zero}\\
    w_0(\rx)&:=&\Im[\tilde\delta_0(\rx)]+\frac{1}{2}
    \sum_{k=1}^{\infty} (-1)^{k}\frac{d^k}{d\rx^k}\,\Im[
    \tilde\delta_k(\rx)].
    \label{w-zero}
    \eea
\item For odd values of $\ell$, $f_\ell\approxN 0$ yields:
    \be
    u_{\ell-}(\rx)\approxN \Re[\tilde\delta_\ell(\rx)]
    ,~~~~~~w_{\ell -}(\rx)\approxN 0,~~~~~~{\rm for~all}~~
    \ell\in \{1,3,5,\cdots\},
    \label{condi-minus}
    \ee
where
    \bea
    u_{\ell-}(\rx)&:=&\frac{1}{2}\sum_{k=1}^{\infty} (-1)^{\ell+k}
    \left[\frac{(\ell+k)!}{\ell !\: k!}\right]\frac{d^k}{d\rx^k}\,
    \Re[\tilde\delta_{\ell+k}(\rx)],
    \label{u-L-minus}\\
    w_{\ell-}(\rx)&:=&\sum_{k=1}^{\infty} (-1)^{\ell+k}
    \left[\frac{(\ell+k)!}{\ell !\: k!}\right]\frac{d^k}{d\rx^k}\,
    \Im[\tilde\delta_{\ell+k}(\rx)].
    \label{fc4}
    \eea
\item For positive even values of $\ell$, $f_\ell\approxN 0$
yields:
    \be
    u_{\ell+}(\rx)\approxN 0,~~~~~~
    w_{\ell +}(\rx)\approxN \Im[\tilde\delta_\ell(\rx)],~~~~~~
    {\rm for~all}~~\ell\in \{2,4,5,\cdots\},
    \label{condi-plus}
    \ee
where
    \bea
    u_{\ell+}(\rx)&:=&\sum_{k=1}^{\infty} (-1)^{\ell+k}
    \left[\frac{(\ell+k)!}{\ell !\: k!}\right]\frac{d^k}{d\rx^k}\,
    \Re[\tilde\delta_{\ell+k}(\rx)],
    \label{u-L-plus}\\
    w_{\ell+}(\rx)&:=&-\frac{1}{2}\sum_{k=1}^{\infty} (-1)^{\ell+k}
    \left[\frac{(\ell+k)!}{\ell !\: k!}\right]\frac{d^k}{d\rx^k}\,
    \Im[\tilde\delta_{\ell+k}(\rx)].
    \label{w-L-plus}
    \eea
\end{itemize}

Unfortunately, the infinite series appearing in the above
relations involve arbitrarily high order derivatives of
$\tilde\delta_\ell$. This reduces their convergence rate
appreciably (compared to the series expansion (\ref{rh-sw=}) for
$\rh$) and amplifies the approximation errors considerably, thus
rendering a numerical verification of (\ref{condi-zero}),
(\ref{condi-plus}), and (\ref{condi-minus}) intractable. However,
the second condition in (\ref{condi-zero}) provides an interesting
explanation for the particular shape of $\Im[\tilde\delta_0]$:
Neglecting all the terms involving $\Im[\tilde\delta_\ell]$ with
$\ell>0$, this condition reads,
    \[\Im[\tilde\delta_0(\rx)]\approx i{\rm v}(\rx)=\left\{
    \begin{array}{ccc}
    -Z&{\rm for}& -1<\rx<0\\
    Z&{\rm for}& 0<\rx<1
    \end{array}\right.\]
This is in remarkable good agreement with graph of
$\Im[\tilde\delta_0(\rx)]$ as depicted in Fig.~\ref{fig5}.

\end{appendix}

\ed

\begin{figure}
\centerline{\epsffile{g.eps}} \caption{graph}
\end{figure}
\newpage

{\small
\begin{thebibliography}{99}
\bibitem{p1} A.\ Mostafazadeh, J.\ Math.\ Phys.\ {\bf 43}, 205 (2002).
\bibitem{p2} A.~Mostafazadeh, J.\ Math.\ Phys.\ {\bf 43}, 2814 (2002).
\bibitem{p3} A.\ Mostafazadeh, J.\ Math.\ Phys., {\bf 43}, 3944
(2002).
\bibitem{bender-98} C.~M.~Bender and S.~Boettcher, Phys.\ Rev.\ Lett.\
{\bf 80}, 5243 (1998).
\bibitem{bender-99} C.~M.~Bender, S.~Boettcher, and P.~N.~Meisenger, J.~Math.\ Phys.\
{\bf 40}, 2201 (1999).
\bibitem{jmp-2004} A.\ Mostafazadeh, J.\ Math.\ Phys., {\bf 45},
932 (2004).
\bibitem{quasi} F.\ G.\ Scholtz, H.\ B.\ Geyer,
and F.\ J.\ W.\ Hahne, Ann.\ Phys.\ {\bf 213}, 74 (1992).
\bibitem{kresh} R.~Kretschmer and L.~Szymanowski, Phys.\ Lett.~A
{\bf 325}, 112 (2004).
\bibitem{ss1} L.\ Solombrino, J.~Math.\ Phys., {\bf 43}, 5439
(2002).
\bibitem{p6} A.~Mostafazadeh, J.\ Math.\ Phys., {\bf 43}, 6343
(2002); Erratum: {\bf 44}, 943 (2003).
\bibitem{ss2} G.\ Scolarici and L.\ Solombrino, J.~Math.\
Phys., {\bf 44}, 4450 (2003).
\bibitem{cjp-2003} A.~Mostafazadeh, Czech J.~Phys.~{\bf 53}, 1079
(2003).
\bibitem{reed-simon} M.\ Reed and B.~Simon, {\em Functional
Analysis,} vol.\ I (Academic Press, San Diego, 1980).
\bibitem{p9} A.~Mostafazadeh, J.\ Math.\ Phys.\ {\bf 44}, 974 (2003)
\bibitem{bbj} C.~M.~Bender, D.~C.~Brody and H.~F.~Jones,
Phys.\ Rev.\ Lett.\ {\bf 89}, 270401 (2002).
\bibitem{jpa-2003} A.~Mostafazadeh, J.~Phys.~A: Math.\ Gen.\
{\bf 36}, 7081 (2003).
\bibitem{critique} A.~Mostafazadeh, quant-ph/0310164.
\bibitem{bbj-ajp} C.~M.~Bender, D.~C.~Brody and H.~F.~Jones,
Am.~J.~Phys.\ {\bf 71}, 1095 (2003).
\bibitem{bbj-erratum} C.~M.~Bender, D.~C.~Brody and H.~F.~Jones,
Phys.\ Rev.\ Lett.\ {\bf 92}, 119902 (2004).
\bibitem{comment} A.~Mostafazadeh, quant-ph/0407070.
\bibitem{cqg} A.~Mostafazadeh, Class.\ Quantum Grav.\ {\bf 20}, 155
(2003).
\bibitem{rqm1} A.~Mostafazadeh, quant-ph/0307059.
\bibitem{bi-ortho}  J.~Wong, J.~Math.\ Phys.\ {\bf 8}, 2039 (1967);\\
F.~H.~M.~Faisal and J.~V.~Moloney, J.~Phys.\ B: At.\ Mol.\ Phys.\
{\bf 14}, 3603 (1981).
\bibitem{p4} A.~Mostafazadeh, Nucl.\ Phys.\ B, {\bf 640}, 419 (2002).
\bibitem{bohm-qm} A.~Bohm, {\em Quantum Mechanics: Foundations and
Applications}, 3rd Edition (Springer, 1993).
\bibitem{z1} M.~Znojil, Phys.\ Lett.~A {\bf 285}, 7 (2001).
\bibitem{bmq} B.~Bagchi, S.~Mallik, and C.~Quesne, Mod.\ Phys.\
Lett,~A {\bf 17}, 1651 (2002).
\bibitem{exceptional} C.~Dembowski, H.~D.~Gr\"af, H.~L.~Harney,
A.~Heine A, W.~D.~Heiss, H.~Rehfeld, and A.~Richter, Phys.\ Rev.\
Lett.\ {\bf 86}, 787 (2001);\\
W.~D.~Heiss and H.~L.~Harney, Eur.\ Phys.~J.\ D {\bf 17}, 149
(2001).
\bibitem{linear-algebra} I.~M.~Gelfand, {\em Lectures on Linear
Algebra} (Dover, New York, 1989).
\bibitem{Horn-Johnson} R.~A.~Horn and C.~R.~Johnson, {\em Matrix
Analysis} (Cambridge University Press, Cambridge, 1985).
\bibitem{Gradshteyn-Ruzhik} I.~S.~Gradshteyn and I.~M.~Ruzhik,
{\em Table of Integrals, Series, and Products} (Academic Press,
San Diego, 1980).
\bibitem{calculus} S.~Lang, {\em A First Course in Calculus}
(Springer, New York, 1998).
\bibitem{other-classical} A.~Nanayakkara, J.~Phys.~A: Math.\ Gen.\
{\bf 37}, 4321 (2004).
\bibitem{wkb} C.~M.~Bender, M.~Berry, P.~N.~Meisinger,
V.~M.~Savage, and M.~Simsek, J.~Phys.~A: Math.\ Gen.\ {\bf 34},
L31 (2001); See also \\
E.~Delabaere and F.~Pham, Phys.\ Lett.~A {\bf 250}, 25 (1998);\\
E.~Delabaere and D.~T.~Trinh, J.~Phys.~A: Math.\ Gen.\ {\bf 33},
8771 (2000).
\bibitem{cjp-2004b} A.~Mostafazadeh, quant-ph/0407213, to appear in
Czech.~J.~Phys.\ (2004).
\bibitem{isham} C.~J.~Isham, {\em Lectures on Quantum Theory}
(Imperial College Press, London, 1995).
\bibitem{rqm} A.~Mostafazadeh and F.~Zamani, quant-ph/0312078.
\bibitem{ap} A.~Mostafazadeh, Ann.~Phys.~(N.Y.) {\bf 309}, 1
(2004).
\bibitem{ahmed} Z.~Ahmed and S.R.~Jain, Phys.~Rev.~E {\bf 67},
045106 (2003); ibid J.\ Phys.\ A {\bf 36}, 3349 (2003).
\bibitem{dynamo} U.~G\"unther and F.~Stefani,
J.~Math.~Phys.\ {\bf 44}, 3097 (2003).
\bibitem{weigert} S.~Weigert, Phys.\ Rev.~A {\bf 68}, 062111
(2003).
\bibitem{bender-2004} C.~M.~Bender, D.~C.~Brody and H.~F.~Jones,
hep-th/0402183 and hep-th/0402011;\\
C.~M.~Bender and H.~F.~Jones, hep-th/0405113.
\bibitem{p62} A.~Mostafazadeh, in preparation.


\end{thebibliography}
}
